\documentclass[11pt,a4paper]{paper}
\usepackage{lmodern}
\usepackage{amsmath,amsfonts,amssymb,amsthm,epsfig,epstopdf,titling,url,array}
\usepackage[T1]{fontenc}
\usepackage{mathrsfs,bbm}
\usepackage[utf8]{inputenc}
\usepackage[english]{babel}
\usepackage{microtype}
\usepackage{changepage,calc}
\usepackage{emptypage}
\usepackage{indentfirst}
\usepackage{booktabs}
\usepackage{tabularx}
\usepackage{array}
\usepackage{graphicx}
\usepackage{subfig}
\usepackage[font={small}]{caption}
\usepackage{listings}
\usepackage[font=small]{quoting}
\usepackage{enumerate}
\usepackage{amsmath,amssymb}
\usepackage{mathtools}
\usepackage{amsthm}
\usepackage[output-decimal-marker={,}]{siunitx}
\usepackage[english]{varioref}
\usepackage{mparhack}
\usepackage{relsize}
\usepackage{bm}
\usepackage{color}
\usepackage{cite}
\numberwithin{equation}{section}

 \newcommand{\ind}{\mathbbm{1}}

 \renewcommand{\tilde}{\widetilde}

\date{}

\begin{document}

\title{A k-means procedure based on a Mahalanobis type distance for clustering multivariate functional data}
\author{\small Andrea Martino$^1$, Andrea Ghiglietti$^2$,
Francesca Ieva$^1$ and Anna Maria
Paganoni$^1$}
\maketitle
\begin{center}
{\small $^1$ MOX - Department of Mathematics, Politecnico di Milano, Milan, Italy\\
\small $^2$ DESP, Universit\`{a} degli Studi di Milano, Milan, Italy}
\end{center}

\abstract{This paper proposes a clustering procedure for samples of multivariate functions in $(L^2(I))^{J}$, with $J\geq1$.
This method is based on a $k$-means algorithm in which the distance between the curves is measured with a metrics
that generalizes the Mahalanobis distance in Hilbert spaces,
considering the correlation and the variability along all the components of the functional data.
The proposed procedure has been studied in simulation and compared with the $k$-means based on other distances
typically adopted for clustering multivariate functional data.
In these simulations, it is shown that the $k$-means algorithm with the generalized Mahalanobis distance provides the best clustering performances,
both in terms of mean and standard deviation of the number of misclassified curves.
Finally, the proposed method has been applied to two real cases studies, concerning ECG signals and growth curves,
where the results obtained in simulation are confirmed and strengthened.
}
\\
\noindent
{\bf Keywords}: Multivariate Functional Data,
Distances in $L^2$,
$k$-means algorithm.

\normalsize
\section{Introduction}
\label{intro}
%
%
%
%
The aim of cluster analysis is to individuate homogenous groups of observations that are realizations of some random process. Clustering is often used as a preliminary step for data exploration, the goal being to identify particular patterns in data that have some convenient interpretation for the user. In particular, $k$-means algorithm is a clustering procedure based on heuristic and geometric procedures.\par
Over the past few decades, in many scientific fields as economics, medicine, engineering, $\ldots$ there has been an increasing interest towards the study of datasets whose number $n$ of statistical units is much smaller than the number $p$ of features recorded for a single statistical unit. \textit{Large p - small n problems} is the term generally used to refer to such situations. A particular case is represented by the situation in which any observed data can be seen as a random function generated by a continuous time stochastic process $X=\{X(t),\: t \in I\}$, lying in a suitable infinite dimensional Hilbert space, typically $L^2(I)$, with $I$ compact interval of $\mathbb{R}$.   \par
\textit{Functional Data Analysis} (FDA) represents the natural framework to develop statistical models and tools which are useful for the study of this kind of data (see, e.g. \cite{Ramsay.et.al.02}, \cite{Ramsay.et.al.05}, \cite{Ferraty.et.al.06}, \cite{Horvath.et.al.12}). As highlighted in this literature, a central role in this context is represented by the Functional Principal Component Analysis (FPCA), which is based on the Karhunen-Lo\`{e}ve (KL) expansion, that decomposes a random function $X(t)$ in a sum of the mean $m(t)$ and a series of orthonormal functions $\varphi_k(t)$, each one multiplied by zero-mean uncorrelated random variables $\sqrt{\lambda}_k Z_k$, where $\{\lambda_k;\, k \geq 1\}$ are the eigenvalues of the covariance operator $V$ of $X$ while $\{\varphi_k;\, k \geq 1\}$ are its eigenfunctions. \par
Despite of the great interest in the FPCA, many inferential procedures adopted in the multivariate PCA have not been extended yet to the functional case. For instance, in the multivariate finite dimensional setting the inference on the mean is typically based on the Mahalanobis distance, since it takes into account the correlation among the variables and it weights the components according to their variability. However, when data belongs to an infinite dimensional space, as $(L^2(I))^J$, the Mahalanobis distance is not well defined and the inference is usually realized by considering only the first $K \in \mathbb{N}$ principal components. Although this approach is widely employed in literature, it is based on a semi-distance that, differently from the Mahalanobis case, does not weight more the components with lower variability.

Clustering functional data can also be a difficult task because of the dimensionality of space the data belong to. The lack of a definition for the probability density of a functional random variable and the difficulty to define distances or make estimates on noisy data are some examples of such difficulties. Different approaches have been proposed along years to address these issues; the most popular one consists again in reducing the infinite dimensional problem to a finite one, approximating the data with elements from some finite dimensional space. Then the usual clustering algorithms for finite dimensional data can be performed. When the goal of the analysis consists in describing the shape of $X(t)$, the first $K$ principal components $\{\varphi_k(t),\,k=1,\ldots,K\}$ usually contain all the information needed to represent the data. Nevertheless, when the goal consists in making inference or classifying curves in different groups, considering a fixed number of components may lead to losing some important information on the distribution of $X(t)$ and hence to providing meaningless results.\par
For these reasons, in this paper we perform a clustering procedure based on a distance that takes into account all the components in $(L^2(I))^J$, with $J\geq1$. This distance was proposed and used in an inferential setting in \cite{Ghiglietti.et.al.14, Ghiglietti.et.al.17},
where it is considered as a generalization of the Mahalanobis distance since it weights the different components according to the correlation and the variability of the functional sample.
The type of clustering procedure we propose to be used with this distance consists in the functional $k$-means algorithm,
which is very popular in the literature of classification in functional data analysis (see, e.g. the $k$-means alignment algorithm in \cite{Sangalli.et.al.10}, the core shape modeling approach in \cite{Boudaoud.et.al.10}, the non-parametric time-synchronized iterative mean updating technique in \cite{Liu.et.al.03} or the simultaneously aligning and cluster K-centres model in \cite{Liu.et.al.09}).
We show, both in simulation and in two applications to real case studies, that the $k$-means algorithm with the generalized Mahalanobis distance provides better clustering performances than the $k$-means based on other distances, that are typically used to deal with multivariate functional data.
Moreover, these good results have been obtained either when the difference between the curves involves their macro-structure or when the difference concerns their micro-structure.
We also discuss how to set the parameter used in the generalized Mahalanobis distance in order to get high clustering performances.\par
The paper is structured as follows. The clustering procedure is presented in Section \ref{clustering_method}, with a short introduction on the generalized Mahalanobis distance. In Section \ref{simulation} we present some results in a simulation setting, both in the univariate and multivariate functional framework, in Subsection \ref{univ} and Subsection \ref{multiv}, respectively. In Section \ref{growth} and Section \ref{ECG} we present some results obtained applying the proposed method to two different real case studies, and finally some concluding remarks are discussed in Section \ref{conclusion}.
All the analysis have been carried out using the software R \cite{RCoreTeam.2016} and the codes are available upon request.

%
%

\section{$\mathbf{k}$-means algorithm with the generalized Mahalanobis distance}
\label{clustering_method}
%
%
The aim of this paper is to develop a proper classification procedure in the multivariate functional framework based on the generalized Mahalanobis distance defined and used in \cite{Ghiglietti.et.al.14, Ghiglietti.et.al.17}. We first recall the definition and the main properties of such distance.\par
Let us consider two realizations $\mathbf{a}$ and $\mathbf{b}$ of a multivariate stochastic process $\mathbf{X}=(X_{1},..,X_{J})^\top$, with $J\geq 1$, $X_{i}\in L^2(I)$ for any $i\in\{1,..,J\}$ and $I$ compact interval of $\mathbb{R}$. The mean $\bm{m}=\mathbb{E}[\bm{X}]$ is defined as a vector of functions in $L^2(I)$ such that $m_l=\mathbb{E}[X_l]$ for any $l \in \{1,..,J\}$, and the covariance kernel $v(s,t)=\mathbb{C}\text{ov}\left[\bm{X}(s), \bm{X}(t)\right]$ is defined as a $J\times J$ matrix of functions such that $v_{l_1l_2}(s,t):=\mathbb{C}\text{ov}\left[X_{l_1}(s),X_{l_2}(t)\right]$ for any $l_1,l_2 \in \{1,\ldots,J\}$. The scalar product between two elements $\mathbf{a}$ and $\mathbf{b}$ of $(L^2(I))^J$ is defined as follows:
\begin{equation*}
\langle \mathbf{a},\mathbf{b}\rangle=\sum_{l=1}^{J}\int_T a_l(t)b_l(t)dt.
\end{equation*}
The eigenvalues $\{\lambda_k;\, k\geq 1\}$ and the eigenfunctions $\{\boldsymbol{\varphi}_{k}=(\varphi_k^{(1)}, \ldots, \varphi_k^{(J)})^\top;\\ k\geq 1\}$ of $v$ are the elements solving
$\langle \mathbf{v}_{l_1\cdot}(t,\cdot),\boldsymbol{\varphi}_k\rangle=\lambda_k\varphi_k^{(l_1)}(t)$ for any $l_1\in\{1,..,J\}$ and $t\in I$, where $\mathbf{v}_{l_1\cdot} = (v_{l_11}, \ldots, v_{l_1 J})$.
Then we can define the generalized Mahalanobis distance as follows:
\begin{equation}
\label{def:dp}
d_p(\mathbf{a},\mathbf{b}) := \sqrt{\sum_{k=1}^{\infty} d_{M,k}^2(\mathbf{a},\mathbf{b}) h_k(p)},
\end{equation}
where $d_{M,k}(\mathbf{a},\mathbf{b})$ indicates the term representing the contribution of the Mahalanobis distance along the $k^{th}$ component, i.e.
\begin{equation*}
d_{M,k}(\mathbf{a},\mathbf{b})= \sqrt{\frac{(\langle \mathbf{a} - \mathbf{b}, \boldsymbol{\varphi}_k\rangle)^2}{\lambda_k}} = \sqrt{\frac{1}{\lambda_k}\Bigg(\sum_{l=1}^{J}\int_T (a_l(t)-b_l(t)) \varphi_k^{(l)}(t)dt\Bigg)^2},
\end{equation*}
and $h_k(p)$ is a sequence of regularizing functions of a suitable real parameter $p>0$. Without loss of generality, throughout all the paper we consider $h_k(p) = \lambda_k/(\lambda_k + 1/p)$, although other choices are possible. For further details on the properties of the $d_p$ distance and the choice of the function $h_k(p)$, see \cite{Ghiglietti.et.al.14, Ghiglietti.et.al.17}.\par
We consider a sample of $n = n_1 + \,\ldots\, + n_k$ realizations $\mathbf{X}_1(t), ...,\mathbf{X}_n(t)$ of $k$ independent stochastic processes in $(L^2(I))^J$. Let $\bar{\mathbf{X}}_n(t) = n^{-1}(\mathbf{X}_1(t) + \ldots + \mathbf{X}_n(t))$ be the empirical mean and then the estimated covariance function is defined as follows:
\begin{equation}
\label{hat_v}
\hat{v}(s,t) := \frac{1}{n-1} \sum_{i=1}^n \big(\mathbf{X}_i(s) - \bar{\mathbf{X}}_n(s)\big)\big(\mathbf{X}_i(t) - \bar{\mathbf{X}}_n(t)\big)^\top,
\end{equation}
from which we can compute the sequences of its eigenfunctions $\{\hat{\boldsymbol{\varphi}}_{k}=(\hat{\varphi}_k^{(1)}, \ldots,\\ \hat{\varphi}_k^{(J)})^\top,\, k\geq 1\}$ and the associated eigenvalues $\{\hat{\lambda}_k;\, k \geq 1\}$. Since in this case the covariance function is computed using $n$ curves, we have $\hat{\lambda}_k = 0$ for all $k\geq n$, and hence the functions $\{\hat{\boldsymbol{\varphi}}_k;\, k \geq n\}$ can be arbitrary chosen such that $\{\hat{\boldsymbol{\varphi}}_k;\, k \geq 1\}$ is an orthonormal basis of $(L^2(I))^J$. \par
The empirical version of the $d_p$ distance based on the covariance estimator $\hat{v}$ can be written as follows:

\begin{equation}\begin{aligned}
\label{def:hat_dp}
\hat{d}^2_p(\textbf{X}_i(t),\textbf{X}_j(t)) &= \sum_{k=1}^{\text{min}\{n-1,T\}} \hat{d}^2_{M,k}(\textbf{X}_i(t), \textbf{X}_j(t)) \hat{h}_k(p) \\
& + \sum_{k=\text{min}\{n-1,T\}+1}^{T} p \Big( \langle \textbf{X}_i(t) - \textbf{X}_j(t), \hat{\boldsymbol{\varphi}}_k \rangle \Big)^2,
\end{aligned}
\end{equation}
where $T$ represents the length of the independent variable grid, while $\hat{d}^2_{M,k}(\cdot,\cdot)$ and $\hat{h}(p)$ represent the estimates of ${d}^2_{M,k}(\cdot,\cdot)$ and $h(p)$ presented in \eqref{def:dp}, using $\{\hat{\lambda}_k;\, k \geq 1\}$ and $\{\hat{\boldsymbol{\varphi}}_k;\, k \geq 1\}$, respectively. Comparing \eqref{def:dp} with \eqref{def:hat_dp}, we can note that, since $\hat{\lambda}_k > 0$ only for $k\leq n-1$ and $(\hat{h}(p)/\hat{\lambda}_k) \rightarrow p$ for $\hat{\lambda}_k \rightarrow 0$, the second term in \eqref{def:hat_dp} makes the expression of $\hat{d}_p$ consistent with the definition of $d_p$ in \eqref{def:dp}.\par
We propose a $k$-means algorithm for an unsupervised classification problem. In \cite{Tarpey.et.al.03} it is possible to find a proper definition of the functional $k$-means procedure and an introduction to its consistency properties.
The functional $k$-means clustering algorithm is an iterative procedure, alternating a step of \textit{cluster assignment}, where all the curves are assigned to a cluster, and a step of \textit{centroid calculation}, where a relevant functional representative (the centroid) for each cluster is identified. More precisely, the algorithm is initialized by fixing the number $k$ of clusters and by randomly selecting a set of $k$ initial centroids $\{\boldsymbol{\chi}_1^{(0)}(t), \ldots , \boldsymbol{\chi}_k^{(0)}(t)\}$ among the curves of the dataset. Given this initial choice, the algorithm iteratively repeats the two basic steps mentioned above. Formally, at the $m^{th}$ iteration of the algorithm, $m\geq 1$, the two following steps are performed:
\begin{description}
\item [\textit{Step 1 (cluster assignment step)}:] each curve is assigned to the cluster with the nearest centroid at the $(m-1)^{th}$ iteration, according to the distance $\hat{d}_p$. Formally, the $m^{th}$ cluster assignment $C_i^{(m)}$ of the $i^{th}$ statistical unit, for $i=1,\ldots,n$, can be written as follows:
\begin{equation*}
C_i^{(m)} := \underset{l=1,\ldots,k}{\operatorname{argmin}}\:\hat{d}_p(\textbf{X}_i(t), \boldsymbol{\chi}_l^{(m-1)}(t));
\end{equation*}
\item [\textit{Step 2 (centroid calculation step)}:] the computation of the centroids  at the $m^{th}$ iteration is performed by solving the optimization problems: for any $l=1,\ldots,k$,
\begin{equation*}
\boldsymbol{\chi}_l^{(m)}(t) := \underset{\boldsymbol{\chi} \in (L^2(I))^J}{\operatorname{argmin}}  \sum_{i:C_i^{(m)} = l} \hat{d}_p(\textbf{X}_i(t),\boldsymbol{\chi}(t))^2,
\end{equation*}
where $C_i^{(m)}$ is the cluster assignment of the $i^{th}$ statistical unit at the $m^{th}$ iteration.
\end{description}
The algorithm stops when the same cluster assignments are obtained at two subsequent iterations, i.e. the set of cluster assignments $\{C_1^{(\bar{m})},\ldots,C_n^{(\bar{m})}\}$ and the set of centroids $\{\boldsymbol{\chi}_1^{(\bar{m})}(t),\ldots,\boldsymbol{\chi}_k^{(\bar{m})}(t)\}$ are considered final solutions of the algorithm if $\bar{m}$ is the minimum integer such that $C_i^{(\bar{m}+1)} \equiv C_i^{(\bar{m})}$ for all $i=1,\ldots,n$.\par
Naturally, the $k$-means procedure does not depend only on the distance adopted in the algorithm, but also on the number of clusters $k$. Since $k$ is typically unknown a priori, we compute the optimal number of clusters $k^*$ via silhouette values and a plot of the final classification, see \cite{Struyf.et.al.97}. In particular, the silhouette plot of a classification consists of a bar plot of the \textit{silhouette values} $s_i$, obtained for each statistical unit $i=1,\ldots,n$ as
\begin{equation*}
s_i := \frac{b_i-a_i}{\text{max}\{a_i,b_i\}},
\end{equation*}
where $a_i$ is the average distance between the $i$th statistical unit and all other ones assigned to the same cluster, whereas
\begin{equation*}
b_i := \underset{l=1,\ldots,k; l\neq C_i}{\operatorname{min}} \frac{\sum_{j:C_j=l} \hat{d}_p(\mathbf{X}_i(t), \mathbf{X}_j(t))}{\#\{j:C_j=l\}}
\end{equation*}
is the minimum average distance of the $i$th statistical unit from another cluster. Clearly $s_i$ always lies between -1 and 1, the former value indicating a misclassified statistical unit while the latter a well classified one.

\section{Simulation Studies}
\label{simulation}
In this section we show some empirical results obtained in simulation to evaluate the performances of the clustering procedure presented in Section \ref{clustering_method}.
\subsection{Simulations in the univariate functional framework}
\label{univ}
\begin{figure}
\centering

\subfloat[][]
   {\includegraphics[width=.5\textwidth, height = 5cm]{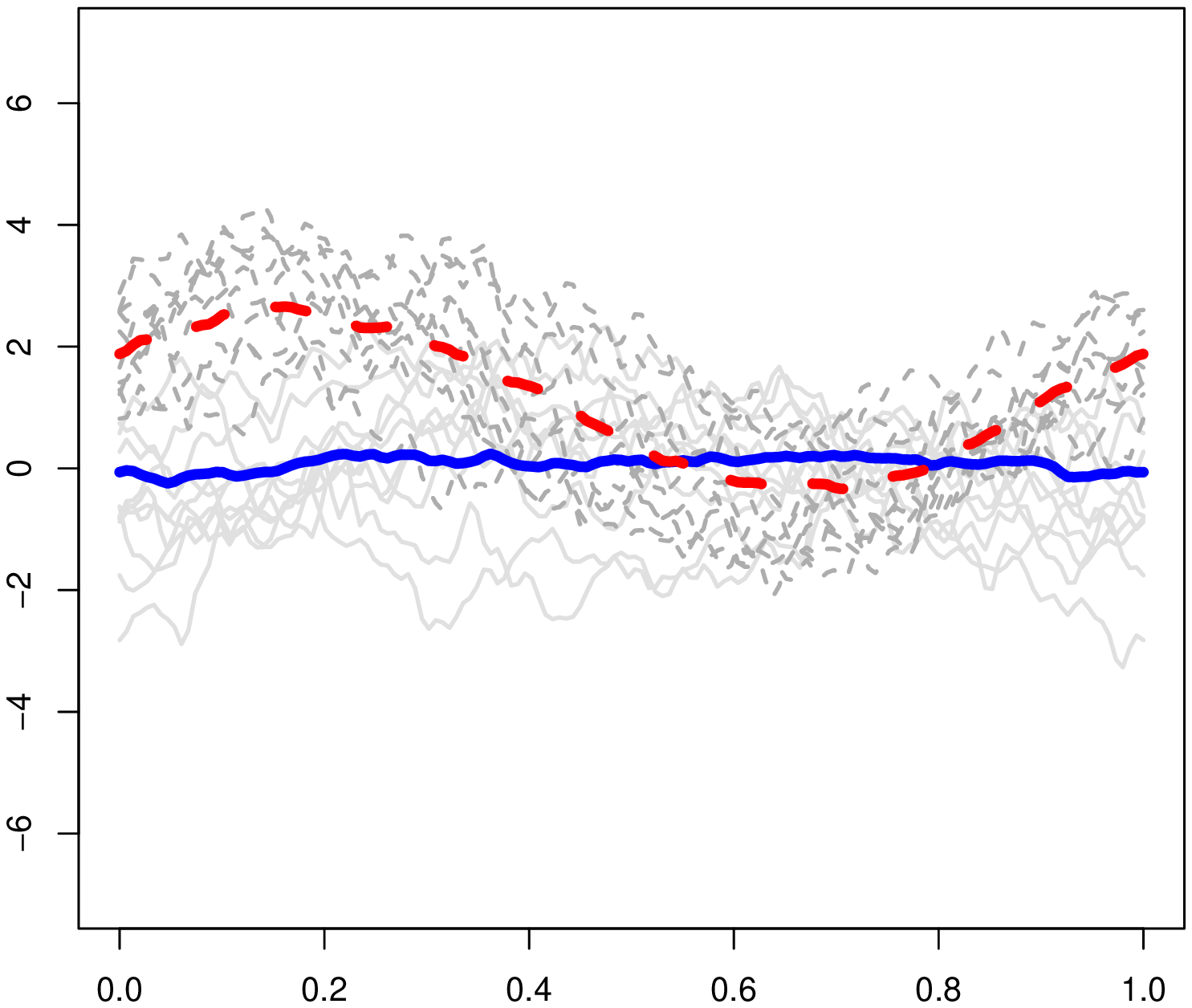}}
\subfloat[][]
   {\includegraphics[width=.5\textwidth, height = 5cm]{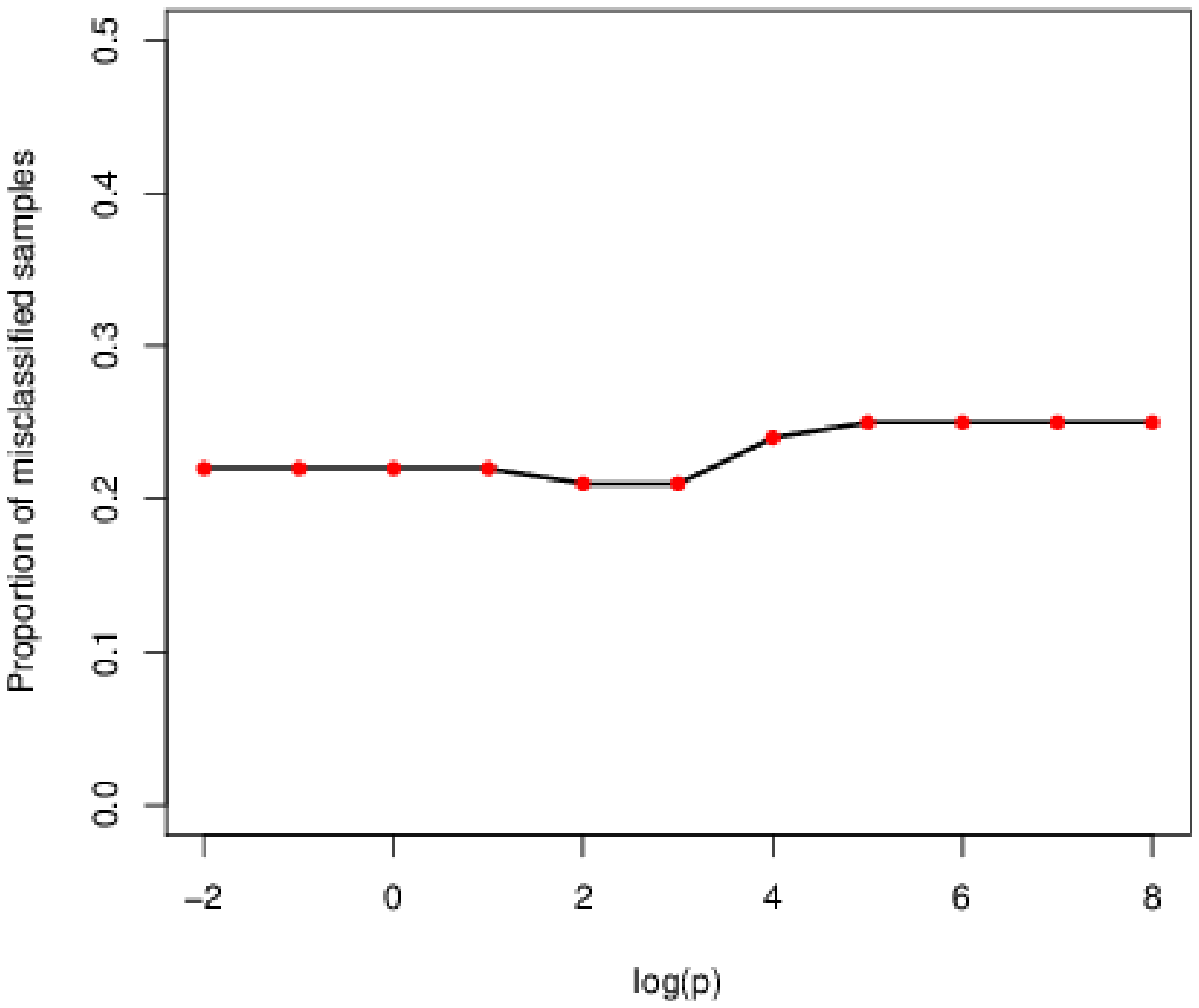}}
\caption[Univariate functional $k$-means procedure for case (i).]{Case (i): $m_2(t) = m_1(t) + \sum_{k=1}^{3} \sqrt{{\rho}_k} {\theta}_k(t).$\\(a) Functional samples $X$ (light grey solid lines) and $Y$ (dark grey dashed lines) along with their sample mean (blue solid line and red dashed line, respectively). \\(b) Proportion of misclassified sample with the functional $k$-means using the $\hat{d}_p$ distance.}
\label{univ1a}
\end{figure}

\begin{figure}
\centering
\includegraphics[width=\textwidth, height = 200px]{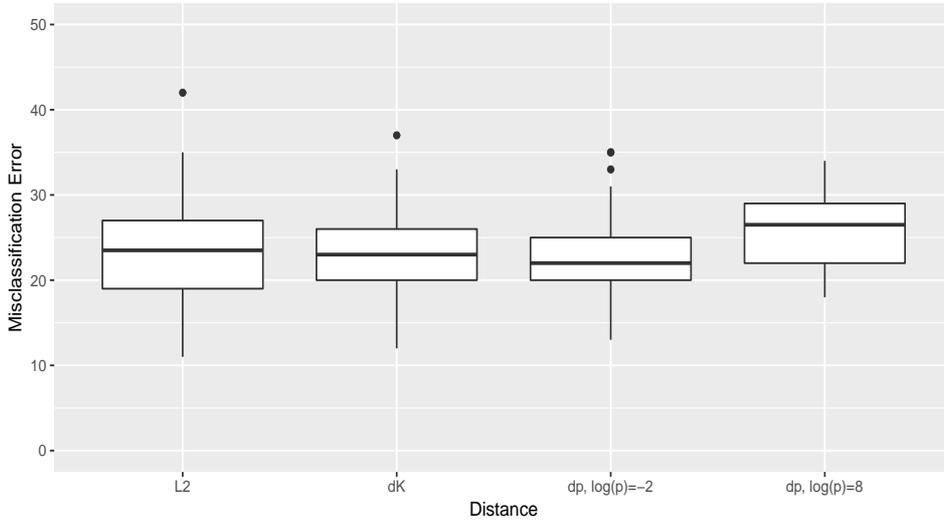}
\caption[]{Boxplot of the number of misclassified curves for case (i) over 50 iterations of the clustering algorithm using  the $L^2$ distance, the truncated Mahalanobis distance $d_M^K$ and the $\hat{d}_p$ distance with log$_{10}(p)=-2$ and log$_{10}(p)=8$, respectively.}
\label{boxplot_i}
\end{figure}

Let us consider two samples of i.i.d. curves $X_1(t), \ldots, X_{n_1}(t)$ and $Y_1(t), \ldots, Y_{n_2}(t)$, generated by independent stochastic processes in $L^2(I)$, with $I$ is a compact interval of $\mathbb{R}$. We generate the sample curves as follows:
\begin{equation*}
X_{i}(t)  = m_1(t)  + \sum_{k=1}^{\tilde{K}} Z_{ki,1} \sqrt{\rho_k} \theta_k(t),  \quad\mbox{for } i = 1, \ldots, n_1,
\end{equation*}
\begin{equation*}
Y_{i}(t)  = m_2(t)  + \sum_{k=1}^{\tilde{K}} Z_{ki,2} \sqrt{\rho_k} \theta_k(t),  \quad\mbox{for } i = 1, \ldots, n_2,
\end{equation*}
where we set:
\begin{enumerate}
\item[(1)] the independent variable grid at $T=150$ equispaced points in $I = [0,1]$;
\item[(2)] $\tilde{K}=100$ components;
\item[(3)] the same sample sizes $n_1 = n_2 = 50$;
\item[(4)] the mean of the first sample $m_1(t) = t (1-t)$, while we  set different values for the mean of the second sample;
\item[(5)] $\{Z_{ki,1};\, k=1,\ldots,\tilde{K} \}$ and $\{Z_{ki,2};\, k=1,\ldots,\tilde{K} \}$ are two collections of independent standard normal variables;
\item[(6)] $\{\rho_k;\, k \geq 1\}$ is a sequence of positive real numbers defined as follows:
\begin{equation*}
\rho_k = \begin{cases}
\frac{1}{k+1} & \text{if $k \in \{1,2,3\},$}\\
\frac{1}{(k+1)^2} & \text{if $k \geq 4;$}
\end{cases}
\end{equation*}
\item[(7)] $ \{ \theta_k;\, k \geq 1\}$ is an orthonormal basis of $L^2(I)$ defined as follows:
\begin{equation*}
\theta_k = \begin{cases}
\ind_{[0,1]}(t) & \text{if $k = 1,$}\\
\sqrt{2} $sin$(k \pi t)\ind_{[0,1]}(t) & \text{if $k \geq 2$, $k$ even,}\\
\sqrt{2} $cos$((k-1) \pi t)\ind_{[0,1]}(t) & \text{if $k \geq 3$, $k$ odd.}
\end{cases}
\end{equation*}
\end{enumerate}
We generate the curves in two different cases:
\begin{enumerate}[(i)]
\item $m_2(t) = m_1(t) + \sum_{k=1}^{3} \sqrt{{\rho}_k} {\theta}_k(t);$
\item $m_2(t) = m_1(t) + \sum_{k=4}^{\tilde{K}} \sqrt{{\rho}_k} {\theta}_k(t).$
\end{enumerate}
We compute the estimated eigenvalues $\{\hat{\lambda}_k;\, k \geq 1\}$ and the associated eigenfunctions $\{\hat{\varphi}_k;\, k \geq 1\}$ from the estimated covariance function $\hat{v}$ as in \eqref{hat_v}, in order to construct the $\hat{d}_p$ distance defined in \eqref{def:hat_dp}.
We compare the performances of the $k$-means based on the $\hat{d}_p$ distance with two competitors: the truncated Mahalanobis semi-distance $d_M^K$ (summing up $K=3$ components, which describe most of the variability) and the $L^2$-distance $d_{L^2}$, as considered in \cite{Horvath.et.al.12}:
\begin{equation}\label{eq:def_competitors}
\begin{aligned}
d_M^K(\mathbf{a},\mathbf{b}) & = \sqrt{\sum_{k=1}^{K} \hat{d}^2_{M,k}(\mathbf{a},\mathbf{b})}\\
& = \sqrt{\sum_{k=1}^K \frac{1}{\hat{\lambda}_k}\Bigg(\sum_{l=1}^{J}\int_T (a_l(t)-b_l(t)) \hat{\varphi}_k^{(l)}(t)dt\Bigg)^2},\\
d_{L^2}(\mathbf{a},\mathbf{b}) & = \|\mathbf{a}-\mathbf{b}\| = \sqrt{ \sum_{l=1}^{J}\int_I (a_l(t)-b_l(t))^2 dt }.
\end{aligned}
\end{equation}
Figure \ref{univ1a} (a) shows the two samples X and Y in case (i), where the two means $m_1(t)$ and $m_2(t)$ differ only along the first three components. Table \ref{tbl:univ1a} shows the results over $M=50$ iterations of the $k$-means algorithm using all the distances mentioned above while Figure \ref{univ1a} (b) shows the proportion of misclassified curves with the $\hat{d}_p$ distance as function of log$_{10}(p)$. Since in case (i) there is a great difference in the macro-structure of the data, the $L^2$-distance $d_{L^2}$ seems to work well, assigning approximately $76\%$ of the data to the right group. For what concerns the other two distances, both the truncated Mahalanobis semi-distance $d_M^K$ and the generalized Mahalanobis distance $\hat{d}_p$ with low values of the parameter $p$ provide quite good results as well. Nevertheless, by looking at Figure \ref{boxplot_i} and Table \ref{tbl:univ1a}, it is possible to note that the $\hat{d}_p$ distance with low values of $p$ gives the best results, both in terms of mean and standard deviation of the number of correctly classified curves.
When the value of $p$ increases,
more elements in $\{\hat{h}_k(p)/\hat{\lambda}_k; k\geq 1\}$, that represents the weights in~\eqref{def:hat_dp}, become close to $1/\hat{\lambda}_k$.
As a consequence, the $\hat{d}_p$ distance gives relevance to a greater number of components,
and so it becomes more similar to the Mahalanobis distance than the $L^2$-distance.
Hence, since in case (i) the curves differ only along three components,
the performances of the clustering procedures get worse.
Indeed, from Figure \ref{univ1a} (b) we can note that the number of misclassified curves increases when $p$ is large,
making the choice of setting a small value of $p$ more appropriate.\par
\begin{table}
\centering
\resizebox{\columnwidth}{!}{\subfloat[][$d_{L^2}$]
{ \begin{tabular}{ccc}
    \toprule
    Cluster & $X$ & $Y$ \\
    \midrule
    1 & \begin{tabular}{@{}c@{}}38.46 \\ \scriptsize(4.6739)\end{tabular} & \begin{tabular}{@{}c@{}}11.54 \\ $\,$ \end{tabular} \\
    2 & \begin{tabular}{@{}c@{}}12.12 \\ $\,$ \end{tabular} & \begin{tabular}{@{}c@{}}37.88 \\ \scriptsize(4.7666)\end{tabular} \\
    \bottomrule
    \multicolumn{3}{c}{Correct classification: .7634}\\
    \bottomrule
  \end{tabular}}
\subfloat[][$d_M^K$]
{ \begin{tabular}{ccc}
    \toprule
    Cluster & $X$ & $Y$ \\
    \midrule
    1 & \begin{tabular}{@{}c@{}}39.04 \\ \scriptsize(3.8701)\end{tabular} & \begin{tabular}{@{}c@{}}10.96 \\ $\,$ \end{tabular} \\
    2 & \begin{tabular}{@{}c@{}}12.26 \\ $\,$ \end{tabular} & \begin{tabular}{@{}c@{}}37.74\\ \scriptsize(4.4895)\end{tabular} \\
   \bottomrule
    \multicolumn{3}{c}{Correct classification: .7678}\\
    \bottomrule
  \end{tabular}}
\subfloat[][$\hat{d}_p$,  log$_{10}(p) = -2$]
{ \begin{tabular}{ccc}
    \toprule
    Cluster & $X$ & $Y$ \\
    \midrule
    1 & \begin{tabular}{@{}c@{}}37.12 \\ \scriptsize(3.6345)\end{tabular} & \begin{tabular}{@{}c@{}}12.88 \\ $\,$ \end{tabular} \\
    2 & \begin{tabular}{@{}c@{}}10.16 \\ $\,$ \end{tabular} & \begin{tabular}{@{}c@{}}39.84 \\ \scriptsize(3.7163)\end{tabular} \\
    \bottomrule
    \multicolumn{3}{c}{Correct classification: \textbf{.7696}}\\
    \bottomrule
  \end{tabular}}
\subfloat[][$\hat{d}_p$, log$_{10}(p)=8$]
{ \begin{tabular}{ccc}
    \toprule
    Cluster & $X$ & $Y$ \\
    \midrule
    1 & \begin{tabular}{@{}c@{}}37.42 \\ \scriptsize(4.7125)\end{tabular} & \begin{tabular}{@{}c@{}}12.58 \\ $\,$ \end{tabular} \\
    2 & \begin{tabular}{@{}c@{}}13.52 \\ $\,$ \end{tabular} & \begin{tabular}{@{}c@{}}36.48 \\ \scriptsize(5.1040)\end{tabular} \\
    \bottomrule
    \multicolumn{3}{c}{Correct classification: .7410}\\
    \bottomrule
  \end{tabular}}}
\caption[Confusion matrices related to the functional $k$-means for the samples ${X}$ and ${Y}$ for case (i)]{Confusion matrices related to the functional $k$-means for the samples ${X}$ and ${Y}$ in case (i).}
\label{tbl:univ1a}
\end{table}

\begin{figure}
\centering
\subfloat[][]
   {\includegraphics[width=.5\textwidth, height = 5cm]{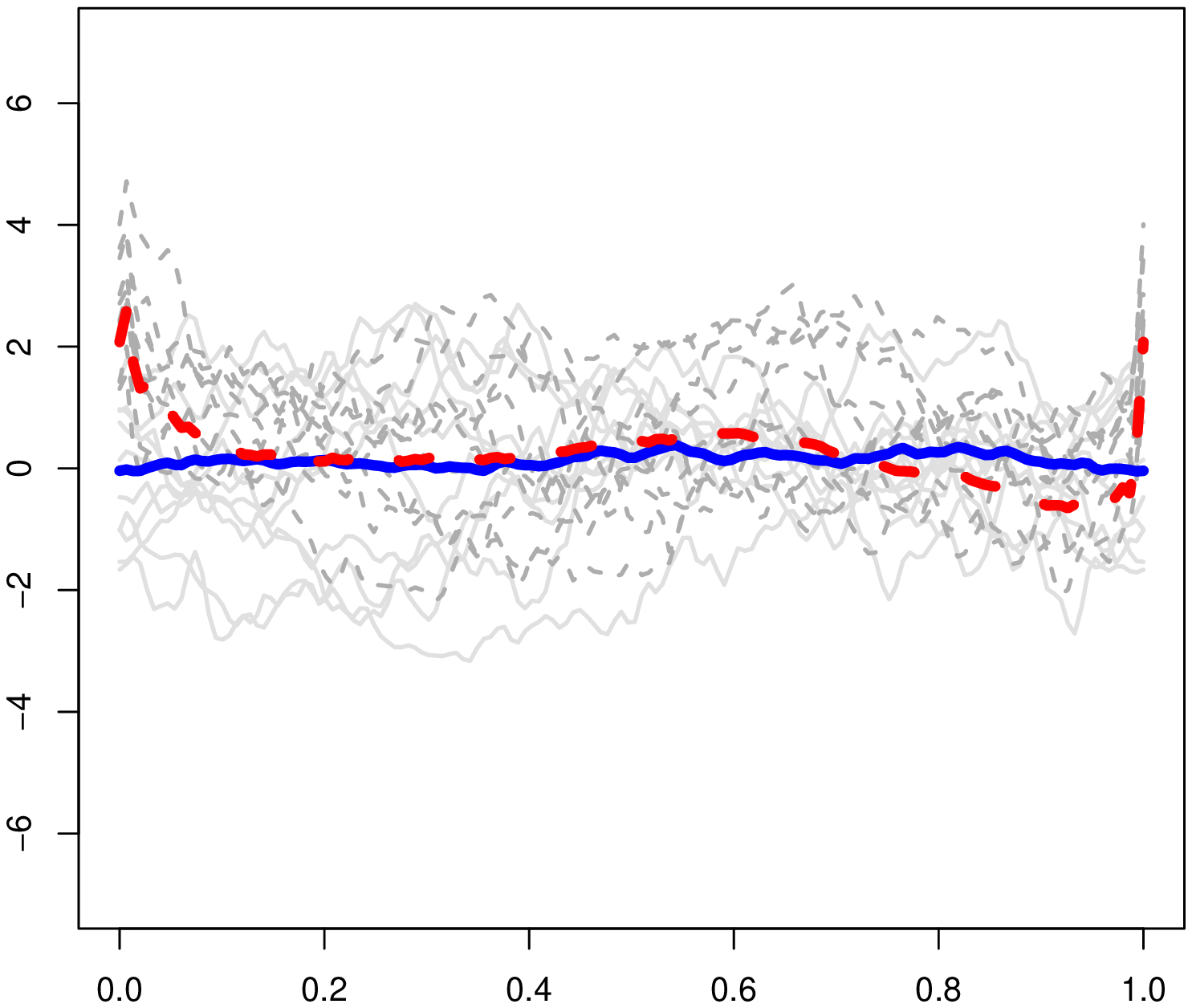}}
\subfloat[][]
   {\includegraphics[width=.5\textwidth, height = 5cm]{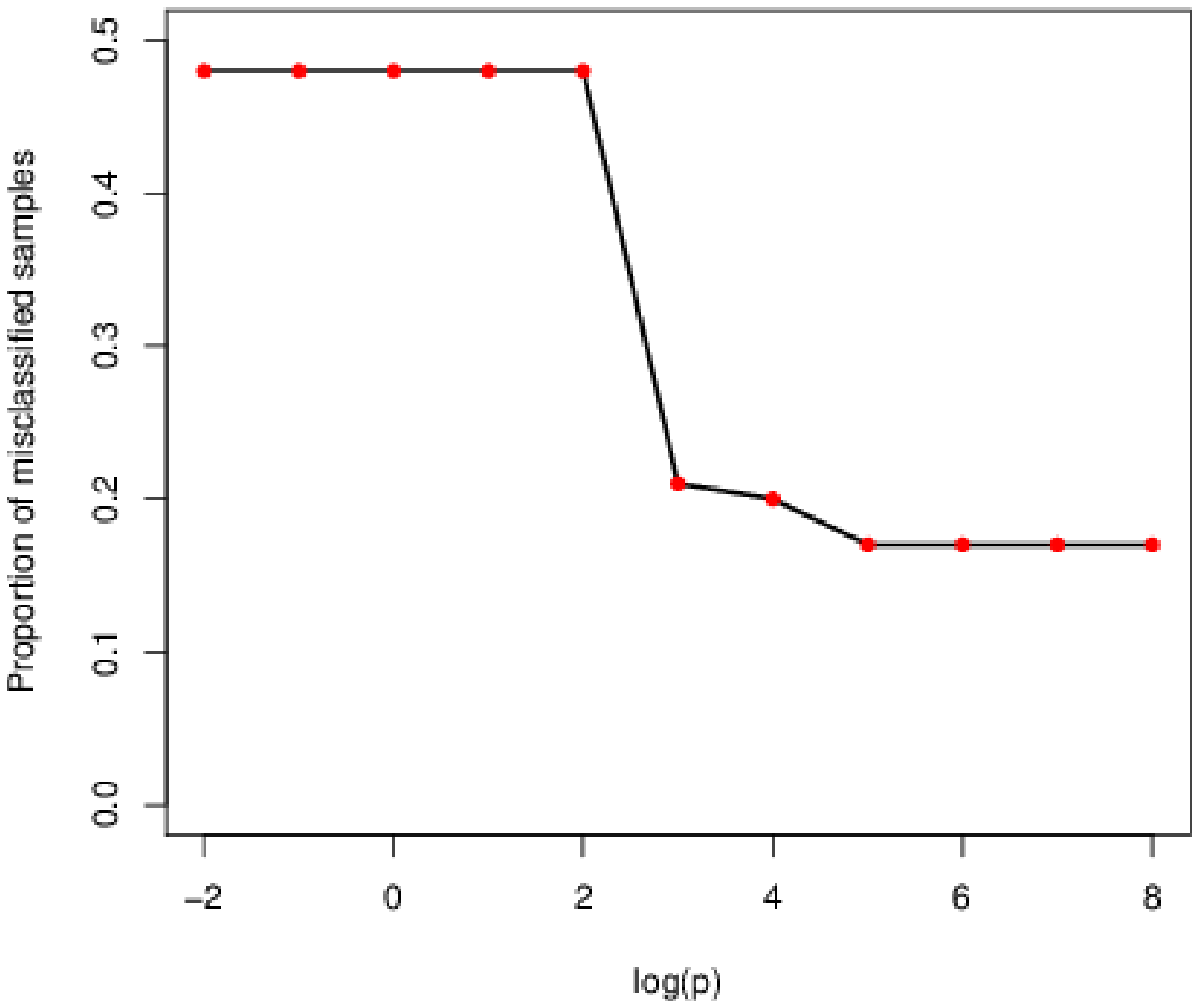}}
\caption[Univariate functional $k$-means procedure for case (ii).]{Case (ii): $m_2(t) = m_1(t) + \sum_{k=4}^{\tilde{K}} \sqrt{{\rho}_k} {\theta}_k(t).$\\(a) Functional samples $X$ (light grey solid lines) and $Y$ (dark grey dashed lines) along with their sample mean (blue solid line and red dashed line, respectively). \\(b) Proportion of misclassified sample with the functional $k$-means using the $\hat{d}_p$ distance.}
\label{univ1c}
\end{figure}

\begin{figure}
\centering
\includegraphics[width=\textwidth, height = 200px]{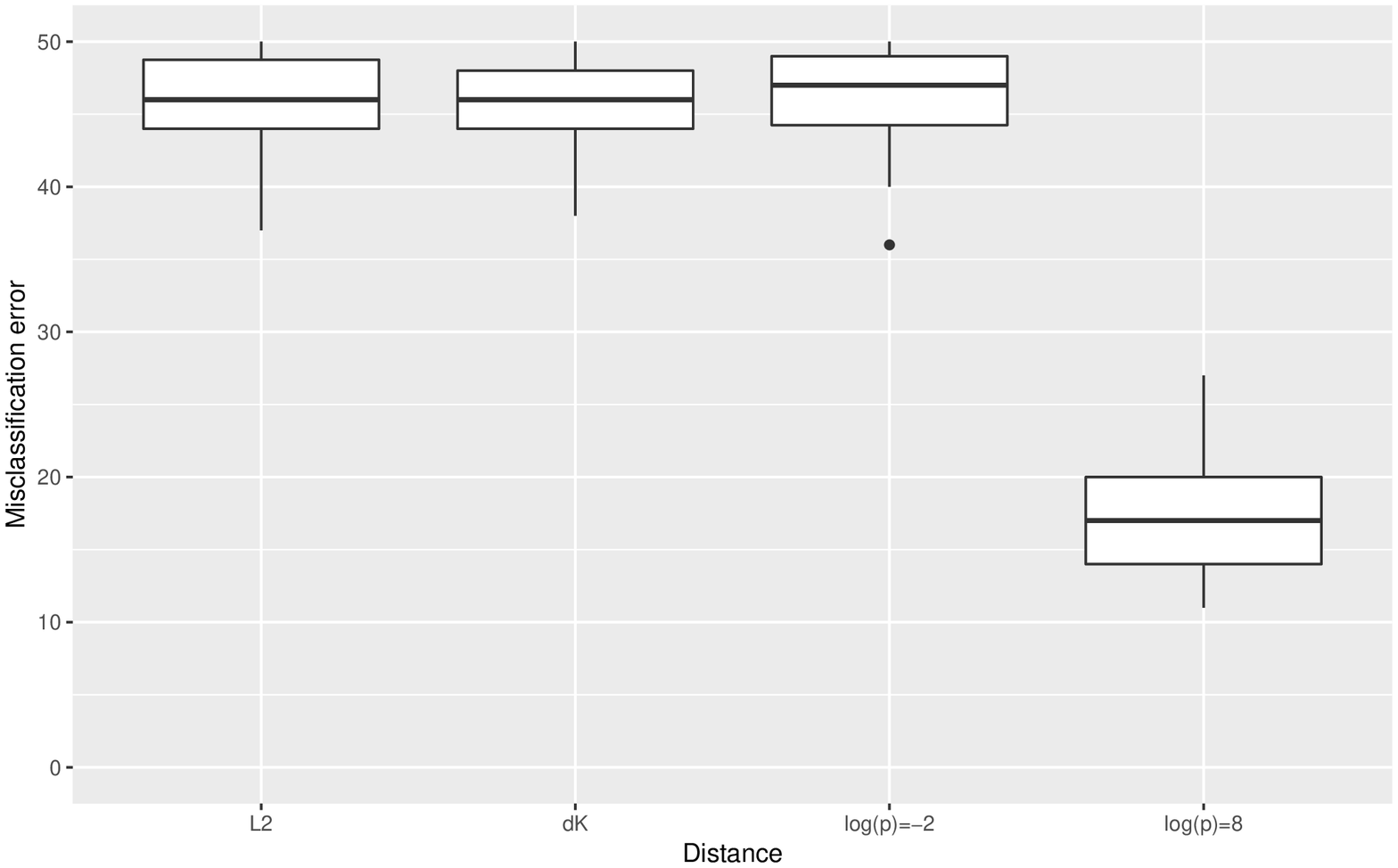}
\caption[]{Boxplot of the number of misclassified curves for case (ii) over 50 iterations of the clustering algorithm using  the $L^2$ distance, the truncated Mahalanobis distance $d_M^K$ and the $\hat{d}_p$ distance with log$_{10}(p)=-2$ and log$_{10}(p)=8$, respectively.}
\label{boxplot_ii}
\end{figure}

The second simulation in the univariate functional framework is given by case (ii), where the two means $m_1(t)$ and $m_2(t)$ differ along all the components except the first three. Figure \ref{univ1c} (a) shows the two samples $X$ and $Y$ in case (ii) and Figure \ref{univ1c} (b) shows the proportion of misclassified curves with the $\hat{d}_p$ distance as function of log$_{10}(p)$. In Table \ref{tbl:univ1c} we can read the results obtained for the $k$-means algorithm over $M=50$ iterations with the respective boxplots in Figure \ref{boxplot_ii}. In this case, the $L^2$-distance and the truncated Mahalanobis semi-distance $d_M^K$ do not work well, since they do not detect the differences between the means; the same occurs for what concerns the $\hat{d}_p$ distance with low values of $p$, because $\hat{h}_k(p) \simeq 0$ for $k \geq 4$ and hence the distance is unable to detect any difference between the curves. As the value of the parameter $p$ increases, more terms in $\{\hat{h}_k(p),\, k\geq 1\}$ become close to one.
As a consequence, the distance takes into account more components and the algorithm works better, assigning more than $80\%$ of the curves to the right group. In this case, the procedure is able to detect the small differences in the micro-structure of the curves due to the components with low variability. \par

\begin{table}
\centering
\resizebox{\columnwidth}{!}{\subfloat[][$d_{L^2}$]
{ \begin{tabular}{ccc}
    \toprule
    Cluster & $X$ & $Y$ \\
    \midrule
    1 & \begin{tabular}{@{}c@{}}26.64 \\ \scriptsize(4.4802)\end{tabular} & \begin{tabular}{@{}c@{}}23.36 \\ $\,$ \end{tabular} \\
    2 & \begin{tabular}{@{}c@{}}22.26 \\ $\,$ \end{tabular} & \begin{tabular}{@{}c@{}}27.74 \\ \scriptsize(3.8376)\end{tabular} \\
    \bottomrule
    \multicolumn{3}{c}{Correct classification: .5438}\\
    \bottomrule
  \end{tabular}}
\subfloat[][$d_M^K$]
{ \begin{tabular}{ccc}
    \toprule
    Cluster & $X$ & $Y$ \\
    \midrule
    1 & \begin{tabular}{@{}c@{}}25.64 \\ \scriptsize(4.4020)\end{tabular} & \begin{tabular}{@{}c@{}}24.36 \\ $\,$ \end{tabular} \\
    2 & \begin{tabular}{@{}c@{}}21.60 \\ $\,$ \end{tabular} & \begin{tabular}{@{}c@{}}28.40 \\ \scriptsize(4.4263)\end{tabular} \\
    \bottomrule
    \multicolumn{3}{c}{Correct classification: .5404}\\
    \bottomrule
  \end{tabular}}

\subfloat[][$\hat{d}_p$,  log$_{10}(p) = -2$]
{ \begin{tabular}{ccc}
    \toprule
    Cluster & $X$ & $Y$ \\
    \midrule
    1 & \begin{tabular}{@{}c@{}}28.18 \\ \scriptsize(4.1634)\end{tabular} & \begin{tabular}{@{}c@{}}21.82 \\ $\,$ \end{tabular} \\
    2 & \begin{tabular}{@{}c@{}}24.30 \\ $\,$ \end{tabular} & \begin{tabular}{@{}c@{}}25.70 \\ \scriptsize(4.4043)\end{tabular} \\
    \bottomrule
    \multicolumn{3}{c}{Correct classification: .5388}\\
    \bottomrule
  \end{tabular}}
\subfloat[][$\hat{d}_p$, log$_{10}(p)=8$]
{ \begin{tabular}{ccc}
    \toprule
    Cluster & $X$ & $Y$ \\
    \midrule
    1 & \begin{tabular}{@{}c@{}}41.80 \\ \scriptsize(3.7796)\end{tabular} & \begin{tabular}{@{}c@{}}8.20 \\ $\,$ \end{tabular} \\
    2 & \begin{tabular}{@{}c@{}}9.30 \\ $\,$ \end{tabular} & \begin{tabular}{@{}c@{}}40.70 \\ \scriptsize(3.4062)\end{tabular} \\
    \bottomrule
    \multicolumn{3}{c}{Correct classification: \textbf{.8250}}\\
    \bottomrule
  \end{tabular}}}
\caption[Confusion matrices related to the functional $k$-means for the samples ${X}$ and ${Y}$ for case (ii)]{Confusion matrices related to the functional $k$-means for the samples ${X}$ and ${Y}$ in case (ii).}
\label{tbl:univ1c}
\end{table}

To conclude, the choice of $p$ should be data-driven. Indeed, if the curves in the sample have a different macro-structure, it is better to set a low value of the parameter $p$, which makes the $\hat{d}_p$ distance similar to the $L^2$-distance. On the contrary, when the curves seem very similar among each other but they differ in the micro-structure, the $L^2$-distance does not work well anymore and the choice of a high value of $p$ is more appropriate.

\subsection{Simulations in the multivariate functional framework}
\label{multiv}
We now extend the results presented in the previous section to the multivariate functional framework. Let us consider two samples of i.i.d. curves, $\mathbf{X}_1(t),\ldots, \mathbf{X}_{n_1}(t)$ and $\mathbf{Y}_1(t),\ldots,\mathbf{Y}_{n_2}(t)$, generated by independent stochastic processes in $(L^2(I))^J$ with $J=2$, where $I$ is a compact interval of $\mathbb{R}$. We generate the sample curves as follows:
\begin{equation*}
\mathbf{X}_{i}(t)  = \mathbf{m}_1(t)  + \sum_{k=1}^{\tilde{K}} \mathbf{Z}_{ki,1} \sqrt{\rho_k} \theta_k(t),  \quad\mbox{for } i = 1, \ldots, n_1,
\end{equation*}
\begin{equation*}
\mathbf{Y}_{i}(t)  = \mathbf{m}_2(t)  + \sum_{k=1}^{\tilde{K}} \mathbf{Z}_{ki,2} \sqrt{\rho_k} \theta_k(t) , \quad\mbox{for } i = 1, \ldots, n_2,
\end{equation*}
where the quantities in the above expressions are the same as those in Section \ref{univ}, except for the following:
\begin{enumerate}
\item[(4new)] the mean of the first sample
\begin{equation*}
     \mathbf{m}_1(t)=\begin{pmatrix}
         t(1-t) \\
         4t^2(1-t) \\
        \end{pmatrix},
  \end{equation*}
while we will set different values for the mean of the second sample;
\item[(5new)] $\{\mathbf{Z}_{ki,1},\, k=1,\ldots,\tilde{K} \}$ and $\{\mathbf{Z}_{ki,2},\, k=1,\ldots,\tilde{K} \}$ are two collections of bivariate normal random variables with mean $\boldsymbol{\mu} = \mathbf{0}$ and covariance matrix
\begin{equation*}
{\Sigma}=\begin{pmatrix}
         1 & 0.5 \\
         0.5 & 1 \\
        \end{pmatrix}.
\end{equation*}
\end{enumerate}
We generate the curves in two different cases:
\begin{enumerate}[(i)]
\setcounter{enumi}{2}
\item $\mathbf{m}_2(t) = \mathbf{m}_1(t) + \mathbf{1} \sum_{k=1}^{3} \sqrt{{\rho}_k} {\theta}_k(t)$;
\item $\mathbf{m}_2(t) = \mathbf{m}_1(t) + \mathbf{1} \sum_{k=4}^{\tilde{K}} \sqrt{{\rho}_k} {\theta}_k(t)$.
\end{enumerate}
\begin{figure}
\centering
\subfloat[][]
   {\includegraphics[width=.25\textwidth, height = 5cm]{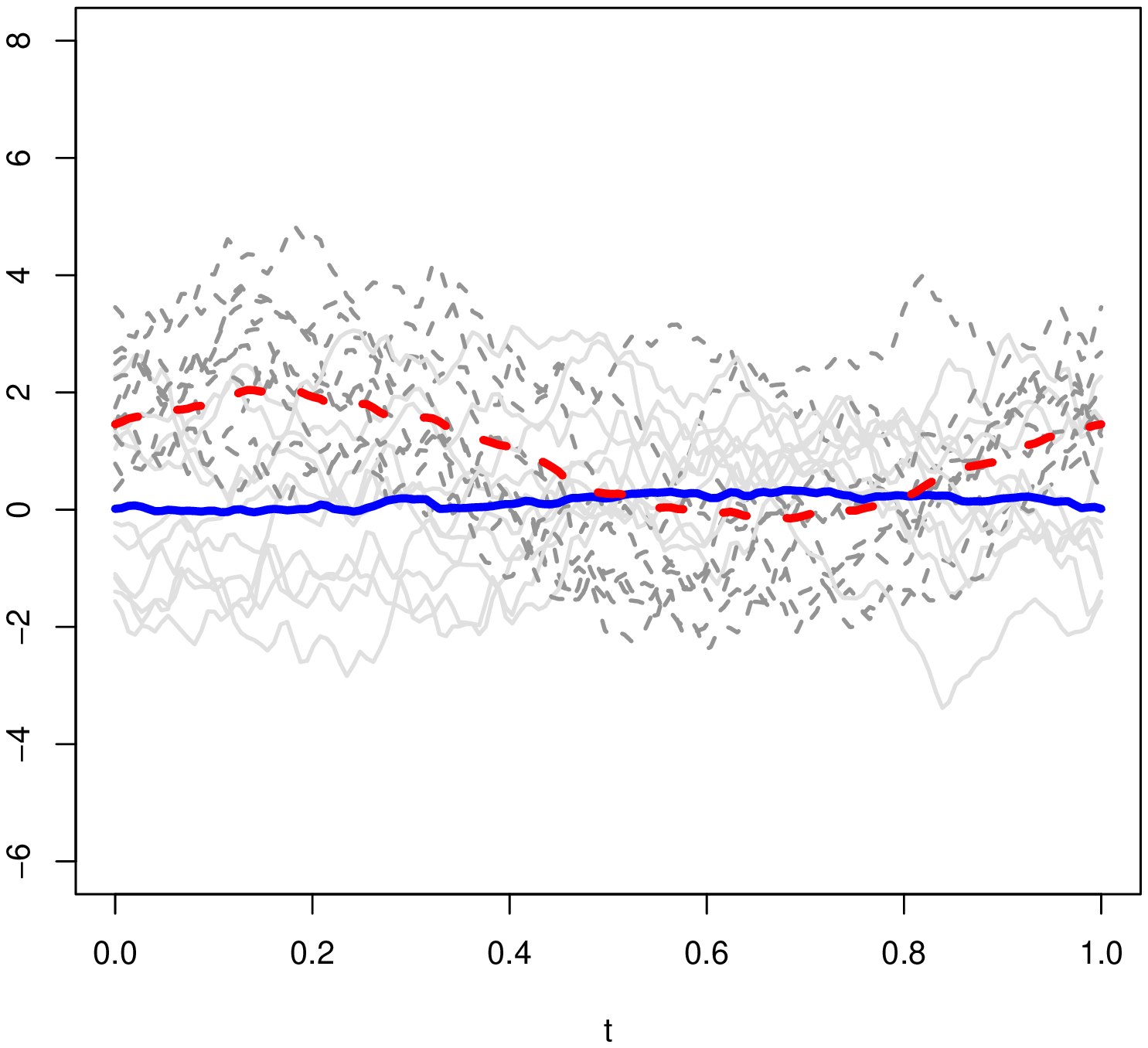}}
   \subfloat[][]
   {\includegraphics[width=.25\textwidth, height = 5cm]{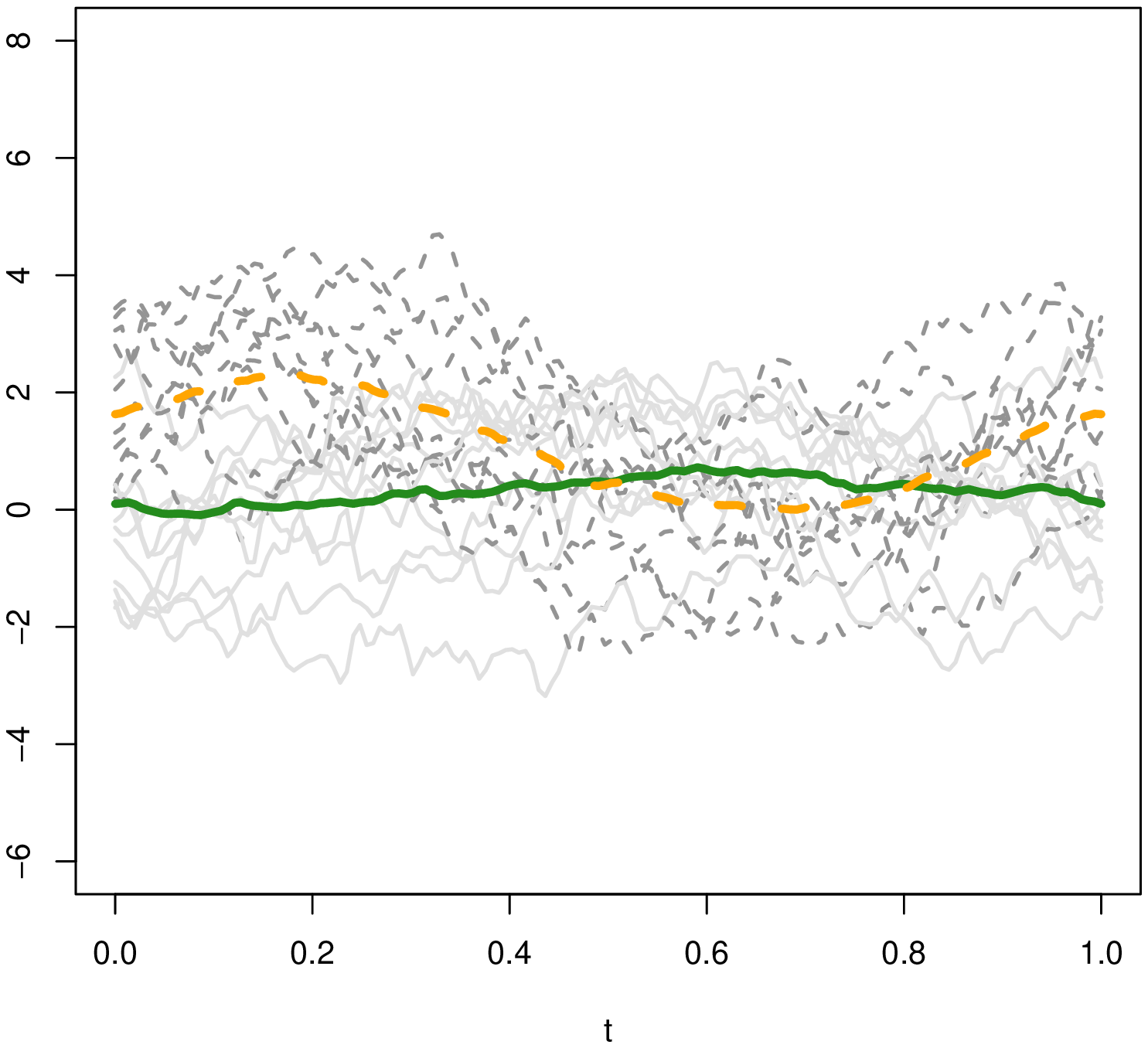}}
\subfloat[][]
   {\includegraphics[width=.5\textwidth, height = 5cm]{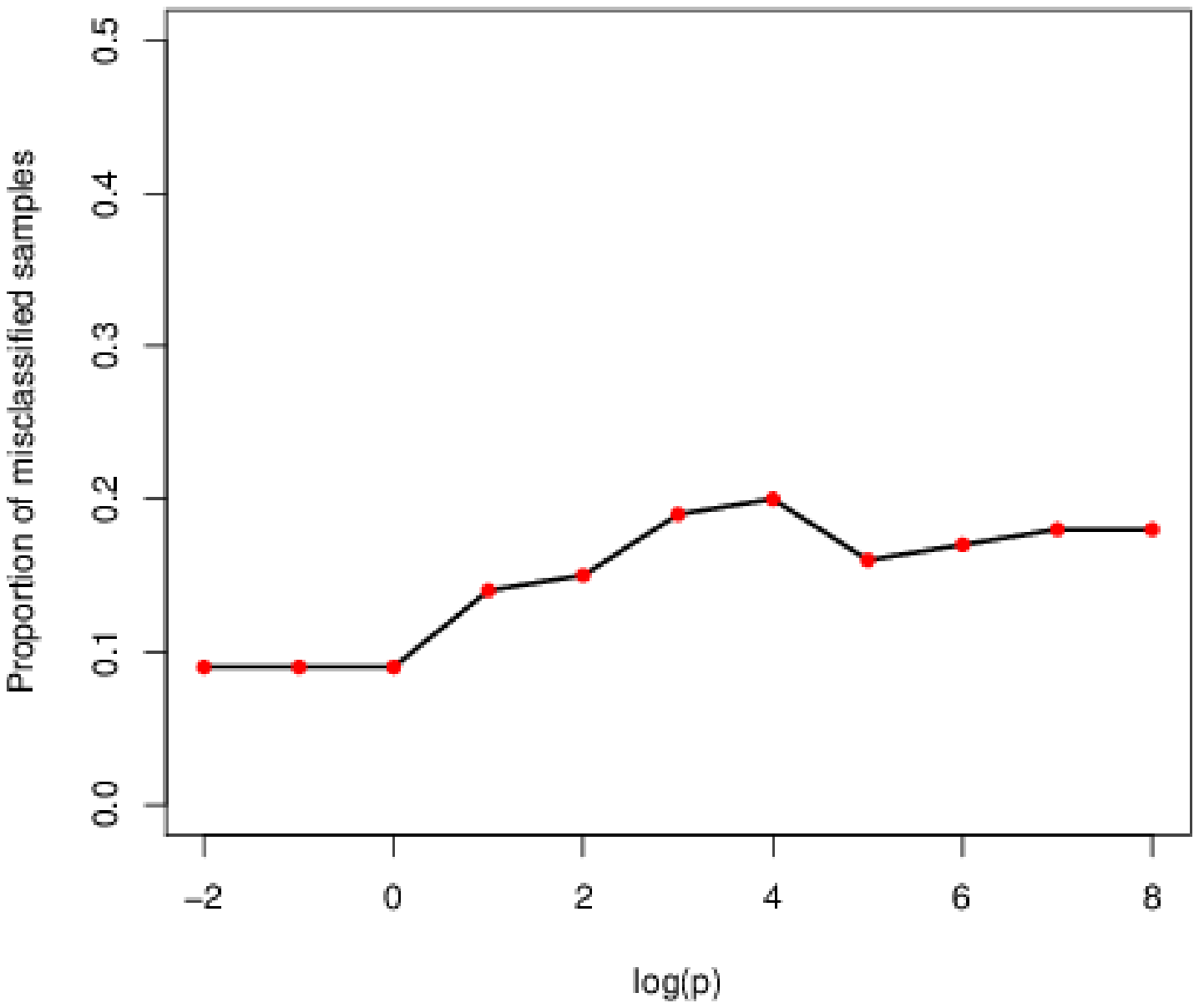}}
\caption[Multivariate functional $k$-means procedure for case (iii).]{Case (iii): $\mathbf{m}_2(t) = \mathbf{m}_1(t) + \mathbf{1} \sum_{k=1}^{3} \sqrt{{\rho}_k} {\theta}_k(t)$.\\(a) First component of the functional samples $X$ (light grey solid lines) and $Y$ (dark grey dashed lines) along with their sample mean (blue solid line and red dashed line, respectively). \\(b) Second component of the functional samples $X$ (light grey solid lines) and $Y$ (dark grey dashed lines) along with their sample mean (green solid line and orange dashed line, respectively). \\ (c) Proportion of misclassified sample with the functional $k$-means using the $\hat{d}_p$ distance.}
\label{fig:multiv2a}
\end{figure}
\begin{table}
\centering
\resizebox{\columnwidth}{!}{\subfloat[][$d_{L^2}$]
{ \begin{tabular}{ccc}
    \toprule
    Cluster & $X$ & $Y$ \\
    \midrule
    1 & \begin{tabular}{@{}c@{}}44.26 \\ \scriptsize(3.4155)\end{tabular} & \begin{tabular}{@{}c@{}}5.74 \\ $\,$ \end{tabular} \\
    2 & \begin{tabular}{@{}c@{}}6.52 \\ $\,$ \end{tabular} & \begin{tabular}{@{}c@{}}43.48 \\ \scriptsize(3.8611)\end{tabular} \\
    \bottomrule
    \multicolumn{3}{c}{Correct classification: .8774}\\
    \bottomrule
  \end{tabular}}
\subfloat[][$d_M^K$]
{ \begin{tabular}{ccc}
    \toprule
    Cluster & $X$ & $Y$ \\
    \midrule
    1 & \begin{tabular}{@{}c@{}}43.50 \\ \scriptsize(3.8611)\end{tabular} & \begin{tabular}{@{}c@{}}6.50 \\ $\,$ \end{tabular} \\
    2 & \begin{tabular}{@{}c@{}}5.96 \\ $\,$ \end{tabular} & \begin{tabular}{@{}c@{}}44.04 \\ \scriptsize(3.1685)\end{tabular} \\
    \bottomrule
    \multicolumn{3}{c}{Correct classification: .8754}\\
    \bottomrule
  \end{tabular}}

\subfloat[][$\hat{d}_p$,  log$_{10}(p) = -2$]
{ \begin{tabular}{ccc}
    \toprule
    Cluster & $X$ & $Y$ \\
    \midrule
    1 & \begin{tabular}{@{}c@{}}43.56 \\ \scriptsize(3.3755)\end{tabular} & \begin{tabular}{@{}c@{}}6.46 \\ $\,$ \end{tabular} \\
    2 & \begin{tabular}{@{}c@{}}5.50 \\ $\,$ \end{tabular} & \begin{tabular}{@{}c@{}}44.50 \\ \scriptsize(3.0921)\end{tabular} \\
    \bottomrule
    \multicolumn{3}{c}{Correct classification: \textbf{.8806}}\\
    \bottomrule
  \end{tabular}}
\subfloat[][$\hat{d}_p$, log$_{10}(p)=8$]
{ \begin{tabular}{ccc}
    \toprule
    Cluster & $X$ & $Y$ \\
    \midrule
    1 & \begin{tabular}{@{}c@{}}41.80 \\ \scriptsize(4.0254)\end{tabular} & \begin{tabular}{@{}c@{}}8.20 \\ $\,$ \end{tabular} \\
    2 & \begin{tabular}{@{}c@{}}8.26 \\ $\,$ \end{tabular} & \begin{tabular}{@{}c@{}}41.74 \\ \scriptsize(3.8269)\end{tabular} \\
    \bottomrule
    \multicolumn{3}{c}{Correct classification: .8354}\\
    \bottomrule
  \end{tabular}}}
\caption[Confusion matrices related to the functional $k$-means for the samples $\mathbf{X}$ and $\mathbf{Y}$ for case (iii)]{Confusion matrices related to the functional $k$-means for the samples $\mathbf{X}$ and $\mathbf{Y}$ in case (iii).}
\label{tbl:multiv2a}
\end{table}
\begin{figure}
\centering
\includegraphics[width=\textwidth, height = 200px]{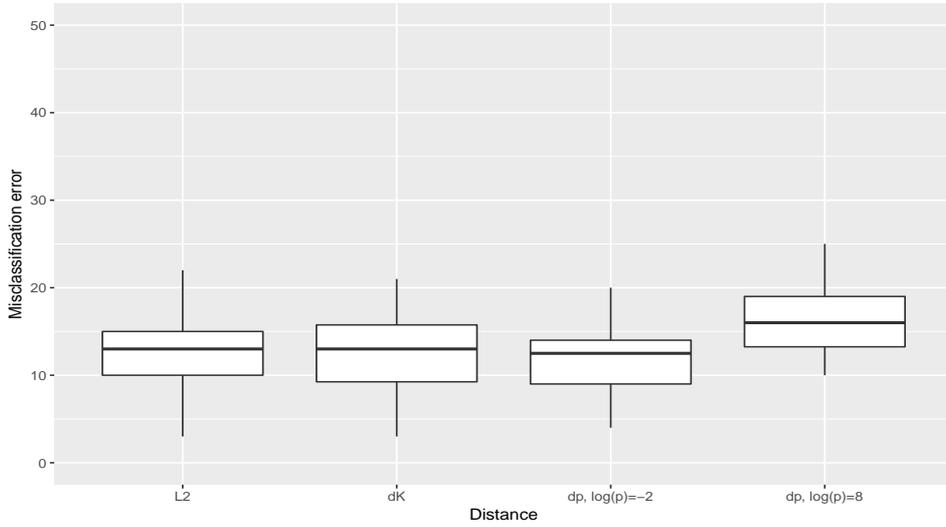}
\caption[]{Boxplot of the number of misclassified curves for case (iii) over 50 iterations of the clustering algorithm using  the $L^2$ distance, the truncated Mahalanobis distance $d_M^K$ and the $\hat{d}_p$ distance with log$_{10}(p)=-2$ and log$_{10}(p)=8$, respectively.}
\label{boxplot_iii}
\end{figure}
We compute the estimated eigenvalues $\{\hat{\lambda}_k;\, k \geq 1\}$ and the associated eigenfunctions  $\{\hat{\boldsymbol{\varphi}}_k(t) = (\hat{\varphi}^{(1)}_k, \hat{\varphi}^{(2)}_k);\, k \geq 1\}$ in order to construct the $\hat{d}_p$ distance as defined in \eqref{def:hat_dp}. 
The truncated Mahalanobis distance $d_M^K$ and the $L^2$-distance $d_{L^2}$ defined in~\eqref{eq:def_competitors}
are again considered as competitors for the $\hat{d}_p$ distance.\par

\begin{figure}
\centering
\subfloat[][]
   {\includegraphics[width=.25\textwidth, height = 5cm]{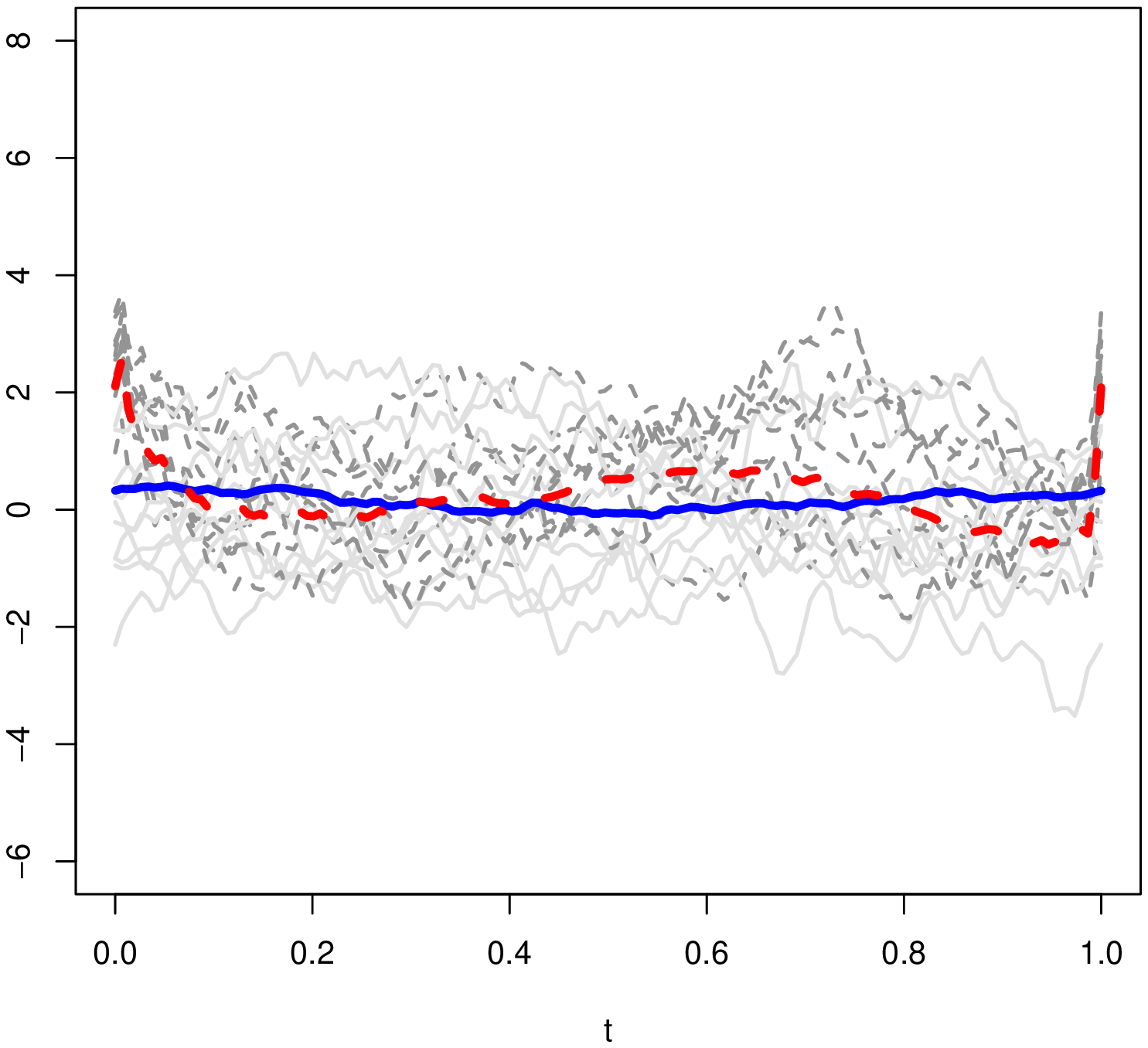}}
   \subfloat[][]
   {\includegraphics[width=.25\textwidth, height = 5cm]{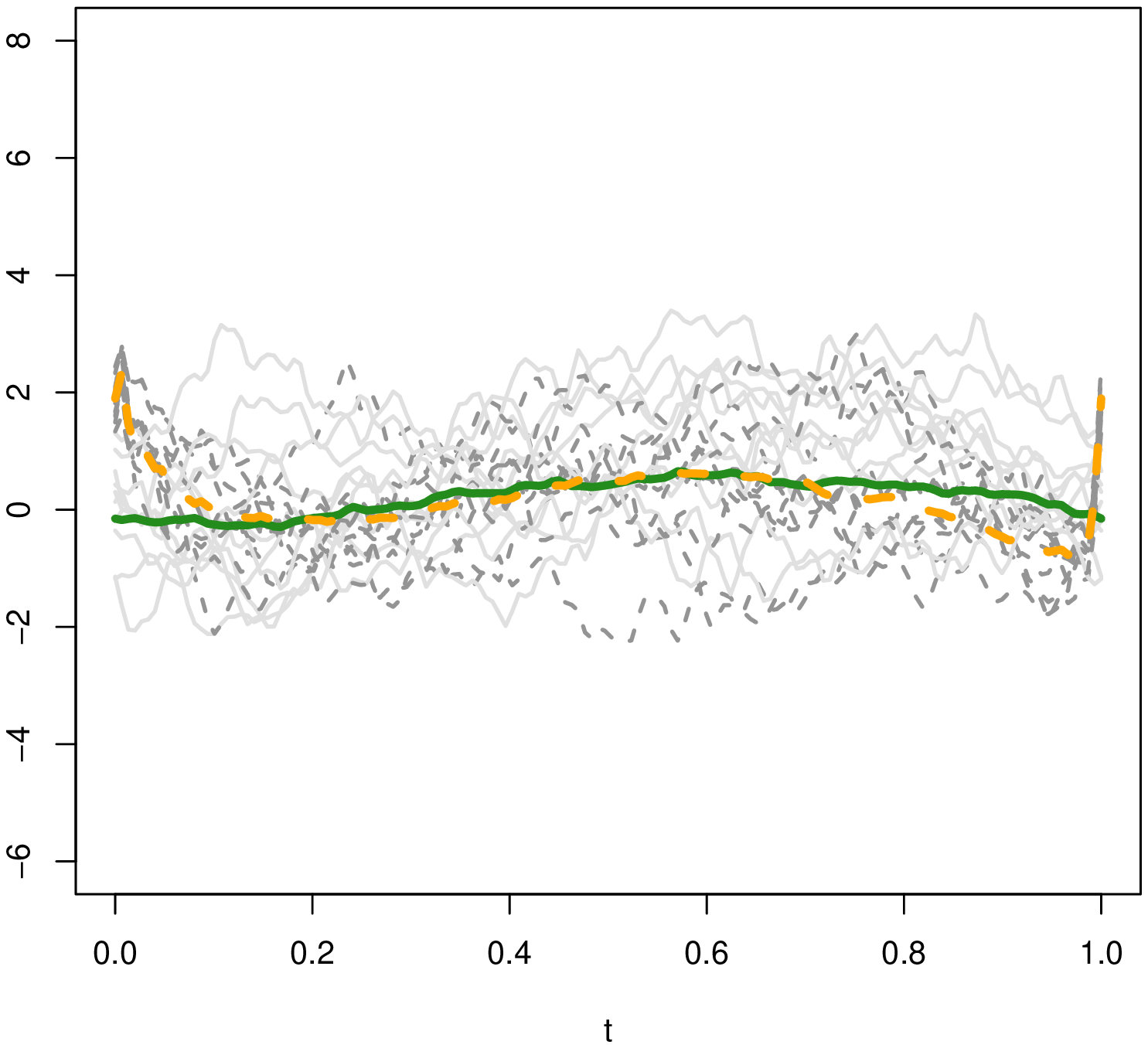}}
\subfloat[][]
   {\includegraphics[width=.5\textwidth, height = 5cm]{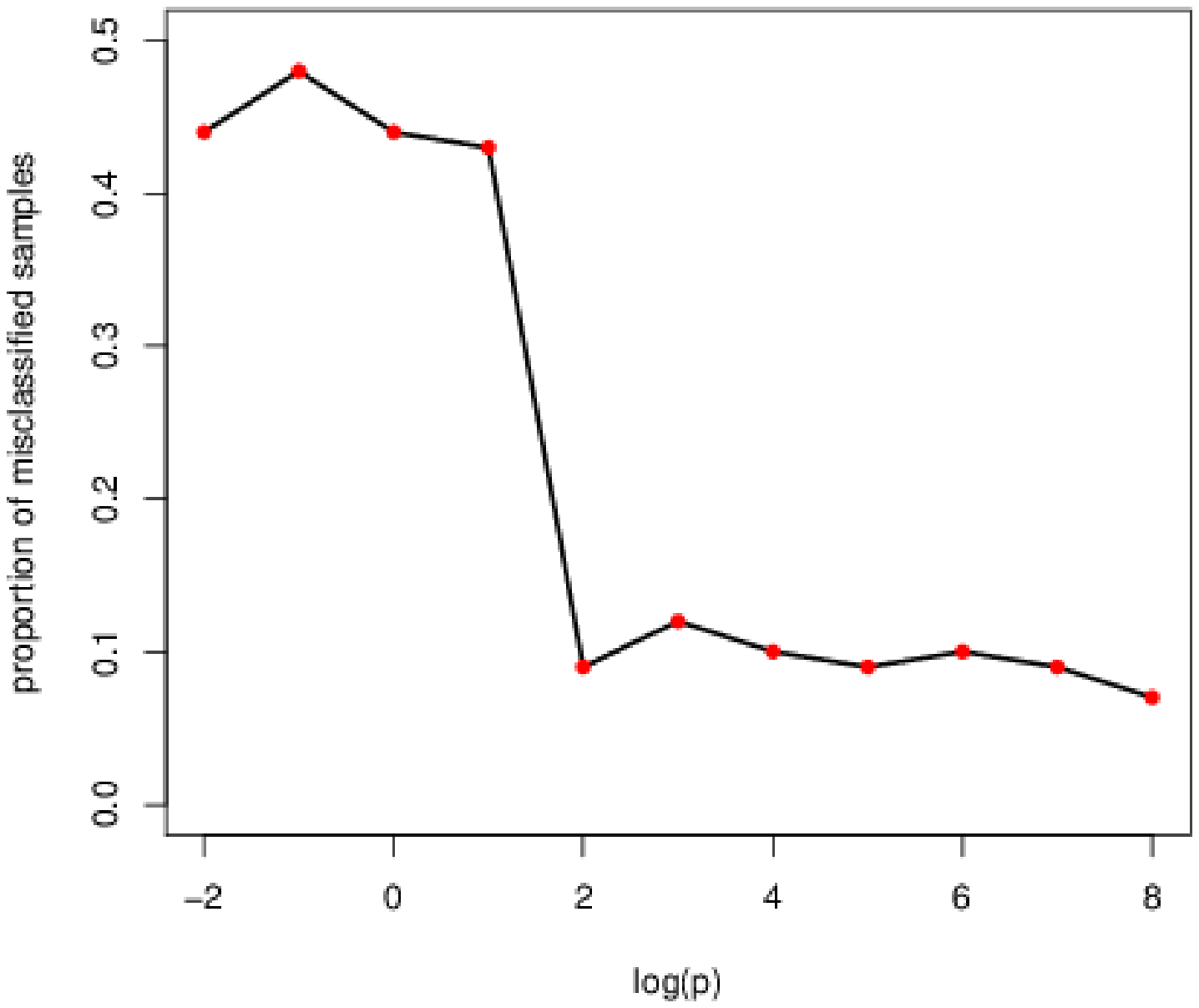}}
\caption[Multivariate functional $k$-means procedure for case (iv).]{Case (iv): $\mathbf{m}_2(t) = \mathbf{m}_1(t) + \mathbf{1} \sum_{k=4}^{\tilde{K}} \sqrt{{\rho}_k} {\theta}_k(t)$.\\(a) First component of the functional samples $X$ (light grey solid lines) and $Y$ (dark grey dashed lines) along with their sample mean (blue solid line and red dashed line, respectively). \\(b) Second component of the functional samples $X$ (light grey solid lines) and $Y$ (dark grey dashed lines) along with their sample mean (green solid line and orange dashed line, respectively). \\ (c) Proportion of misclassified sample with the functional $k$-means using the $\hat{d}_p$ distance.}
\label{fig:multiv2c}
\end{figure}

\begin{table}
\centering
\resizebox{\columnwidth}{!}{\subfloat[][$d_{L^2}$]
{ \begin{tabular}{ccc}
    \toprule
    Cluster & $X$ & $Y$ \\
    \midrule
    1 & \begin{tabular}{@{}c@{}}27.90 \\ \scriptsize(3.7972)\end{tabular} & \begin{tabular}{@{}c@{}}22.10 \\ $\,$ \end{tabular} \\
    2 & \begin{tabular}{@{}c@{}}22.86 \\ $\,$ \end{tabular} & \begin{tabular}{@{}c@{}}27.14 \\ \scriptsize(4.4401)\end{tabular} \\
    \bottomrule
    \multicolumn{3}{c}{Correct classification: .5504}\\
    \bottomrule
  \end{tabular}}
\subfloat[][$d_M^K$]
{ \begin{tabular}{ccc}
    \toprule
    Cluster & $X$ & $Y$ \\
    \midrule
    1 & \begin{tabular}{@{}c@{}}27.24 \\ \scriptsize(4.3685)\end{tabular} & \begin{tabular}{@{}c@{}}22.76 \\ $\,$ \end{tabular} \\
    2 & \begin{tabular}{@{}c@{}}22.14 \\ $\,$ \end{tabular} & \begin{tabular}{@{}c@{}}27.86 \\ \scriptsize(4.1058)\end{tabular} \\
    \bottomrule
    \multicolumn{3}{c}{Correct classification: .5510}\\
    \bottomrule
  \end{tabular}}

\subfloat[][$\hat{d}_p$,  log$_{10}(p) = -2$]
{ \begin{tabular}{ccc}
    \toprule
    Cluster & $X$ & $Y$ \\
    \midrule
    1 & \begin{tabular}{@{}c@{}}27.46\\ \scriptsize(4.6957)\end{tabular} & \begin{tabular}{@{}c@{}}22.54 \\ $\,$ \end{tabular} \\
    2 & \begin{tabular}{@{}c@{}}22.78 \\ $\,$ \end{tabular} & \begin{tabular}{@{}c@{}}27.22 \\ \scriptsize(4.5638)\end{tabular} \\
    \bottomrule
    \multicolumn{3}{c}{Correct classification: .5468}\\
    \bottomrule
  \end{tabular}}
\subfloat[][$\hat{d}_p$, log$_{10}(p)=8$]
{ \begin{tabular}{ccc}
    \toprule
    Cluster & $X$ & $Y$ \\
    \midrule
    1 & \begin{tabular}{@{}c@{}}46.24 \\ \scriptsize(2.1339)\end{tabular} & \begin{tabular}{@{}c@{}}3.76 \\ $\,$ \end{tabular} \\
    2 & \begin{tabular}{@{}c@{}}4.12 \\ $\,$ \end{tabular} & \begin{tabular}{@{}c@{}}45.88 \\ \scriptsize(2.2373)\end{tabular} \\
    \bottomrule
    \multicolumn{3}{c}{Correct classification: \textbf{.9212}}\\
    \bottomrule
  \end{tabular}}}
\caption[Confusion matrices related to the functional $k$-means for the samples $\mathbf{X}$ and $\mathbf{Y}$ for case (iv)]{Confusion matrices related to the functional $k$-means for the samples $\mathbf{X}$ and $\mathbf{Y}$ in case (iv).}
\label{tbl:multiv2c}
\end{table}
\begin{figure}
\centering
\includegraphics[width=\textwidth, height = 200px]{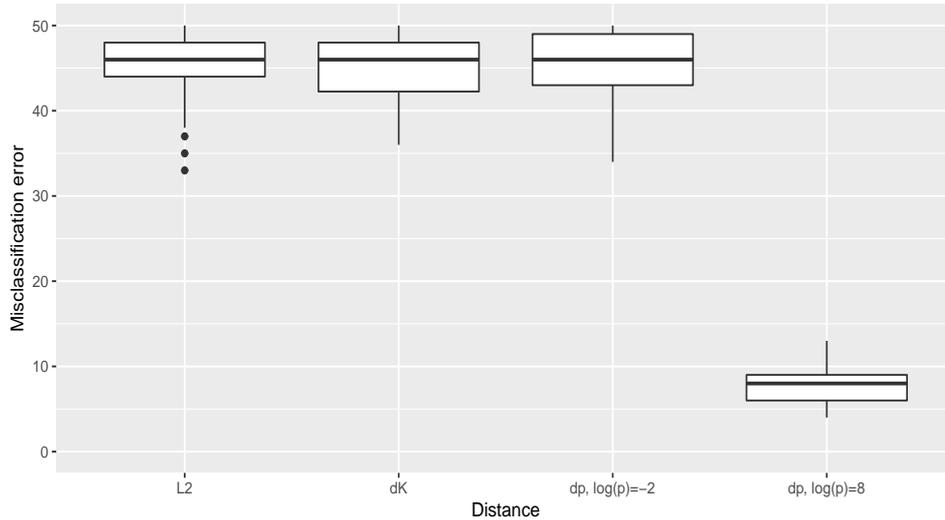}
\caption[]{Boxplot of the number of misclassified curves for case (iv) over 50 iterations of the clustering algorithm using  the $L^2$ distance, the truncated Mahalanobis distance $d_M^K$ and the $\hat{d}_p$ distance with log$_{10}(p)=-2$ and log$_{10}(p)=8$, respectively.}
\label{boxplot_iv}
\end{figure}

Figures \ref{fig:multiv2a} (a-b) show the samples $\mathbf{X}$ and $\mathbf{Y}$ in case (iii), where the means of the two samples differ only along the first three components, while Figure \ref{fig:multiv2a} (c) shows the proportion of misclassified curves using the $\hat{d}_p$ distance as function of log$_{10}(p)$. In Figure \ref{boxplot_iii} and Table \ref{tbl:multiv2a} we can see the results obtained with the three distances over $M = 50$ iterations. The results obtained in this multivariate functional framework confirm and strengthen those obtained in the univariate framework.
Indeed in case (iii), where the difference between the means involves only the components associated with most of the variability,
the $L^2$-distance works quite well, assigning more than 85\% of the curves to the right group. For the other two distances, both the $d_M^K$ distance and the $\hat{d}_p$ distance with low values of $p$ have similar performance, even though the latter works better both in terms of mean and standard deviation of the number of correctly classified curves. Setting a high value of $p$ is not a good choice, since so doing the $\hat{d}_p$ distance considers relevant many components while the curves differs only along three of them.\par
Finally we consider case (iv), where the two means differ along all the components except the first three. Figures \ref{fig:multiv2c} (a-b) show the new samples $\mathbf{X}$ and $\mathbf{Y}$ and Figure \ref{fig:multiv2c} (c) shows the proportion of misclassified curves with the $\hat{d}_p$ distance as function of log$_{10}(p)$. For reasons analogous to those explained in the univariate functional framework, when the value of $p$ is low the $k$-means does not work well and the results are as bad as for the $d_{L^2}$ and $d_M^K$ distances. However, when the value of $p$ increases, the procedure with the $\hat{d}_p$ distance provides very good results (see Figure \ref{fig:multiv2c} (c)), since it takes into account more components of the functional data. Moreover, as it is shown in Figure \ref{boxplot_iv} and Table \ref{tbl:multiv2c}, when we set a high value of $p$, the performances improve considerably also in terms of standard deviation of the number of the correctly classified curves.\par
Therefore, we have shown that all the results obtained in the univariate functional framework also hold in the multivariate functional framework.

\section{Case study I: Growth dataset}
\label{growth}
In this section we apply the clustering procedure proposed in this paper to the Berkeley Growth Study dataset, available in the \texttt{fda} package \cite{fda}, which contains the heights (in cm) of 93 children, measured quarterly from 1 to 2 years, annually from 2 to 8 years and biannually from 8 to 18 years. In the dataset, each function is a univariate curve ($J=1$) defined on a grid of length $T=31$. Out of the 93 children, 39 are boys while 54 are girls, so the aim of the analysis is to point out some differences among them. \par
\begin{figure}
\centering
\subfloat[][]
   {\includegraphics[width=.5\textwidth, height = 5cm]{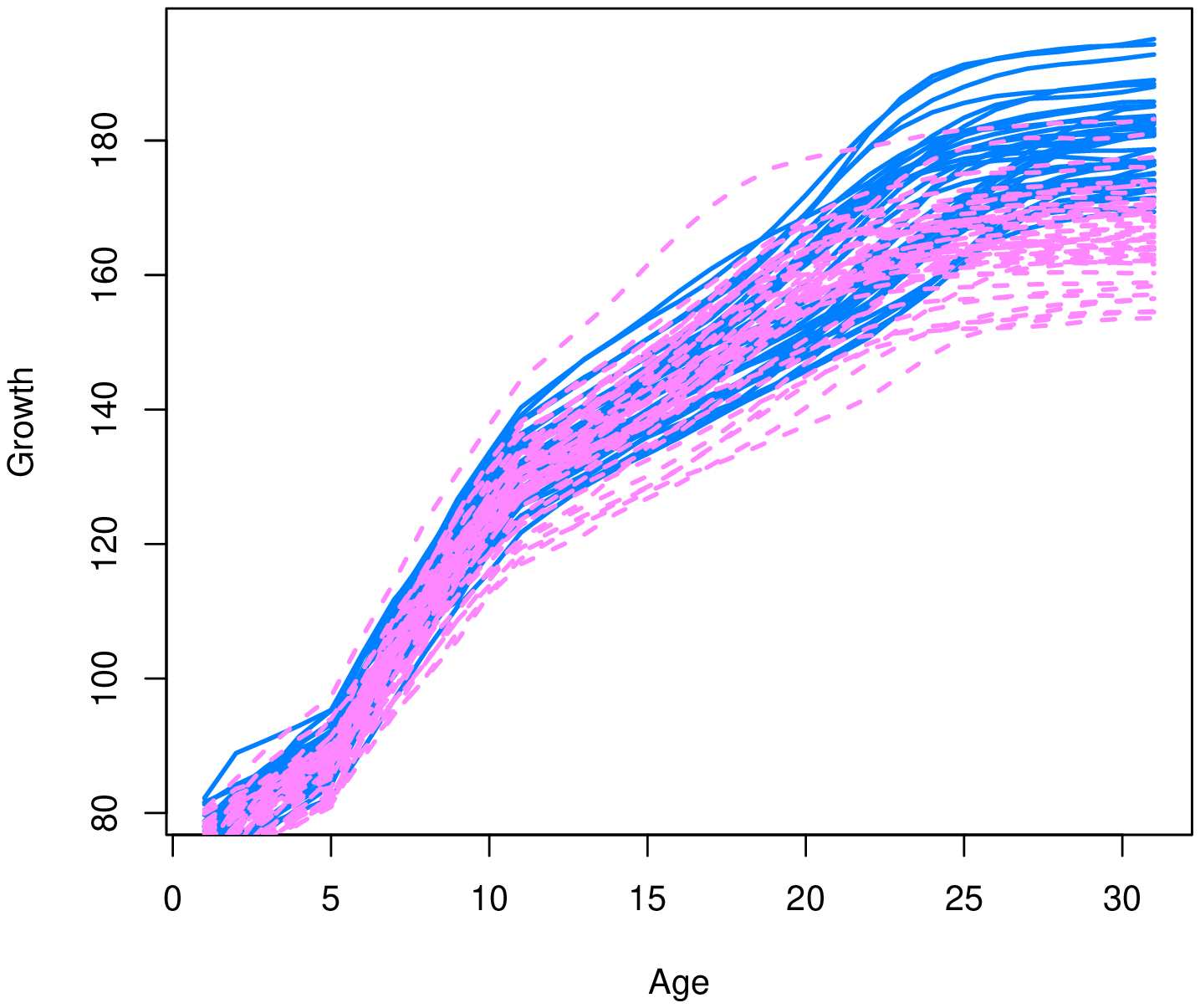}}
\subfloat[][]
   {\includegraphics[width=.5\textwidth, height = 5cm]{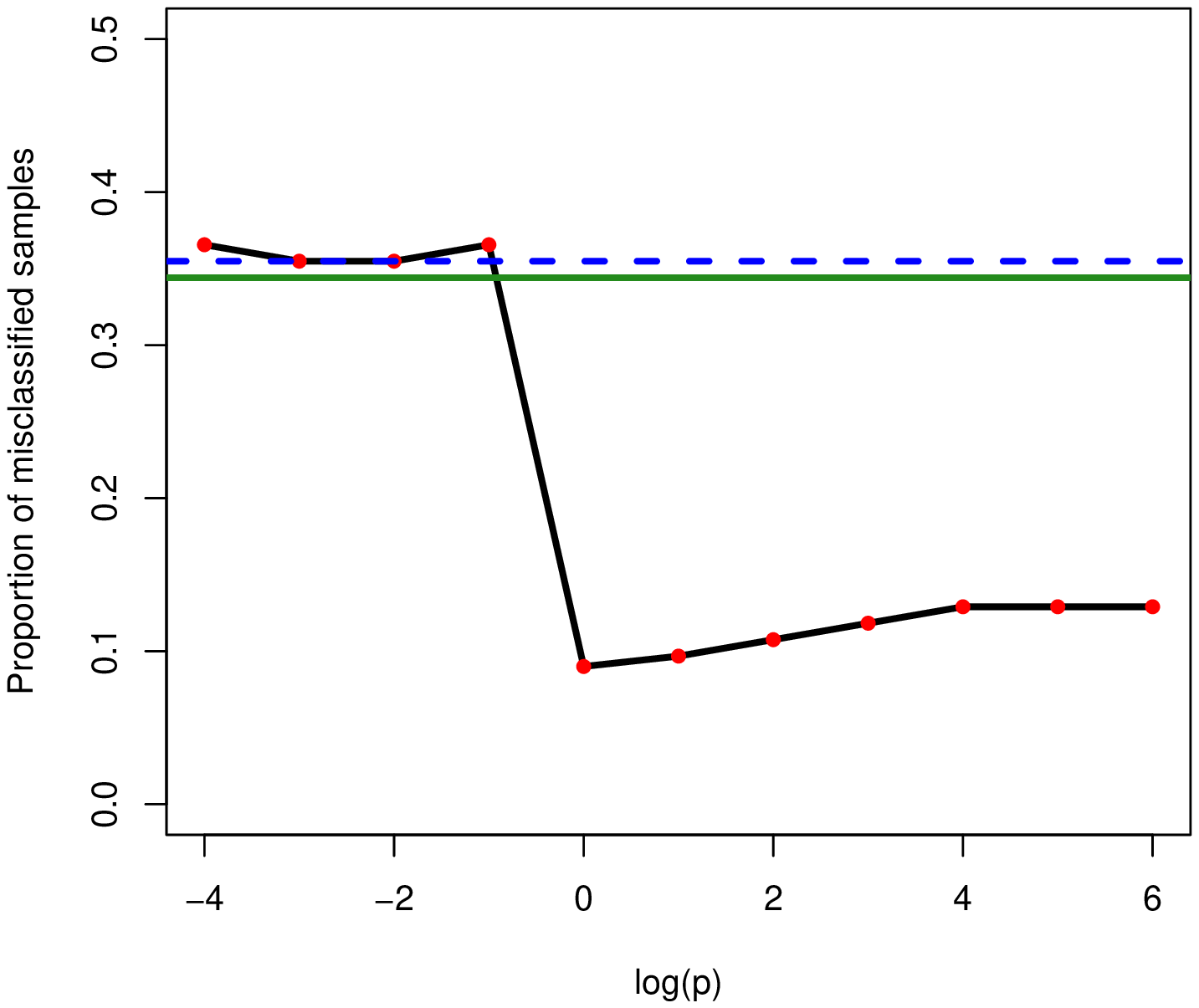}}
\caption[Proportion of misclassified samples with the functional $k$-means for the Growth dataset using the $L^2$ distance, the $d_M^K$ distance and the $\hat{d}_p$ distance.]{Growth dataset. \\ (a) Functional samples for the boys (solid blue lines) and for the girls (dotted pink lines). \\ (b) Proportion of misclassified samples with the functional $k$-means using the $L^2$ distance (blue dashed line), the truncated version of the Mahalanobis distance (green line) and the $\hat{d}_p$ distance (black line).}
\label{fig:growth}
\end{figure}
\begin{figure}
\centering
\subfloat[][]
   {\includegraphics[width=.33\textwidth, height = 3cm]{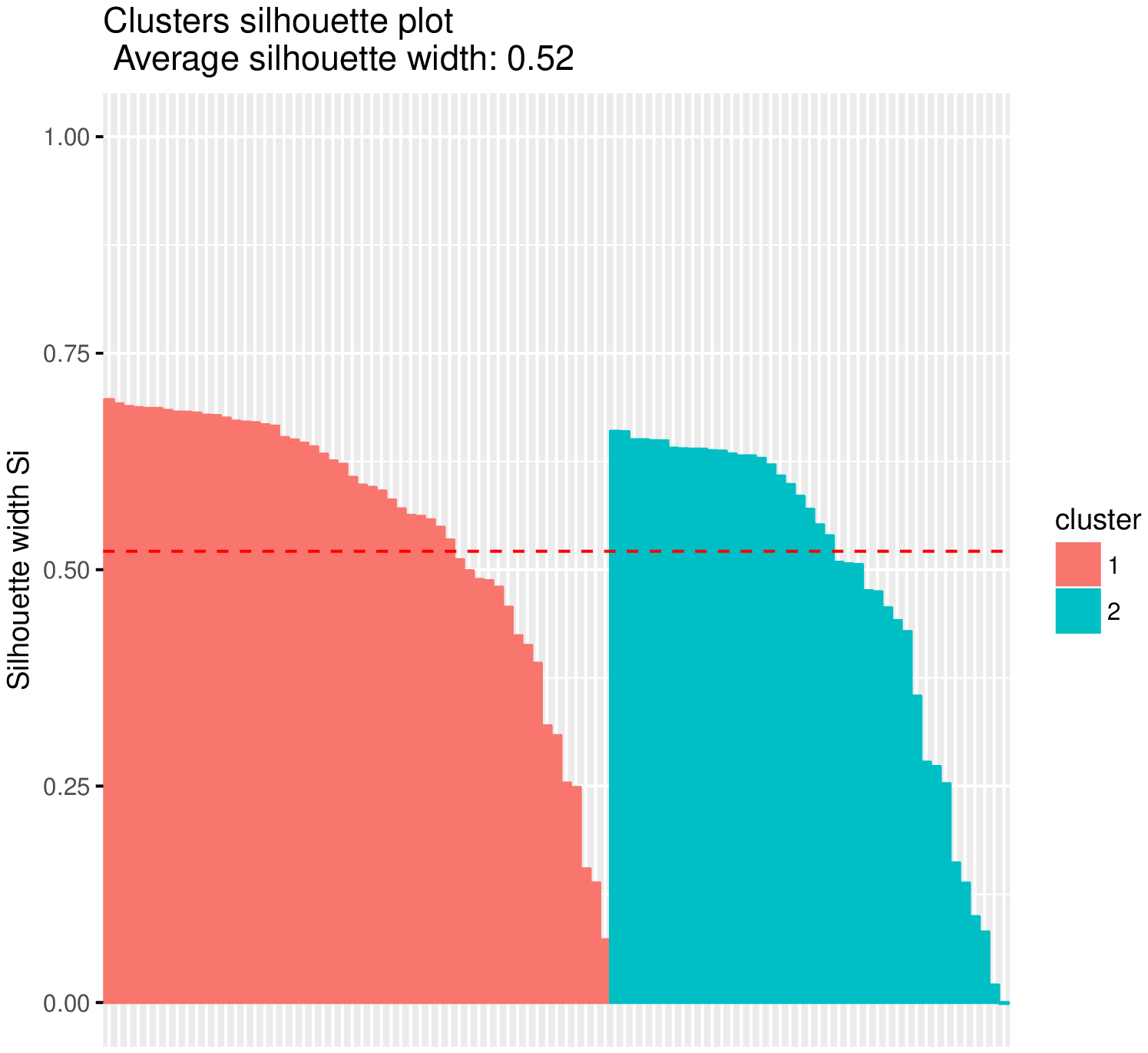}}
\subfloat[][]
   {\includegraphics[width=.33\textwidth, height = 3cm]{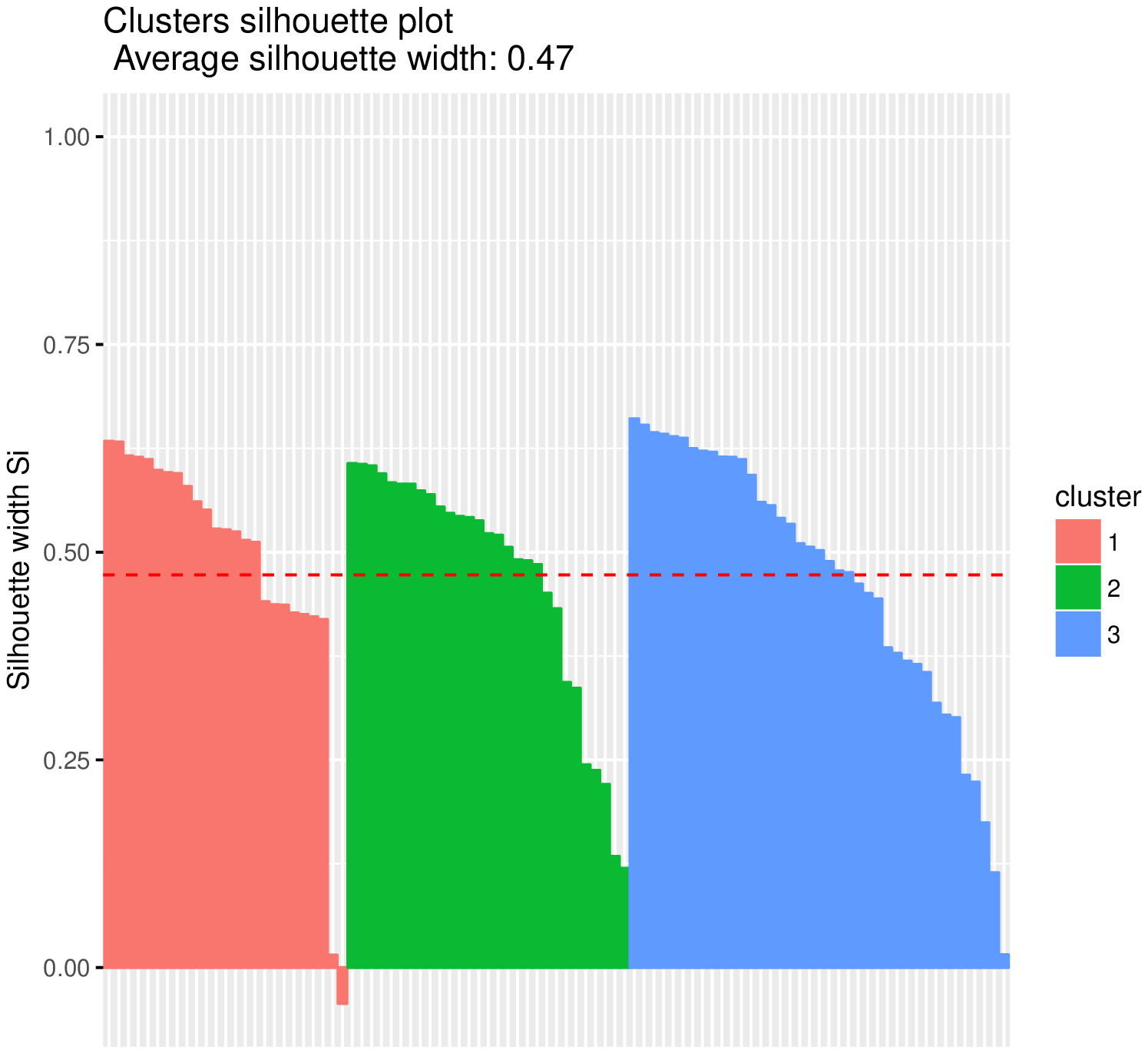}}
   \\
\subfloat[][]
   {\includegraphics[width=.33\textwidth, height = 3cm]{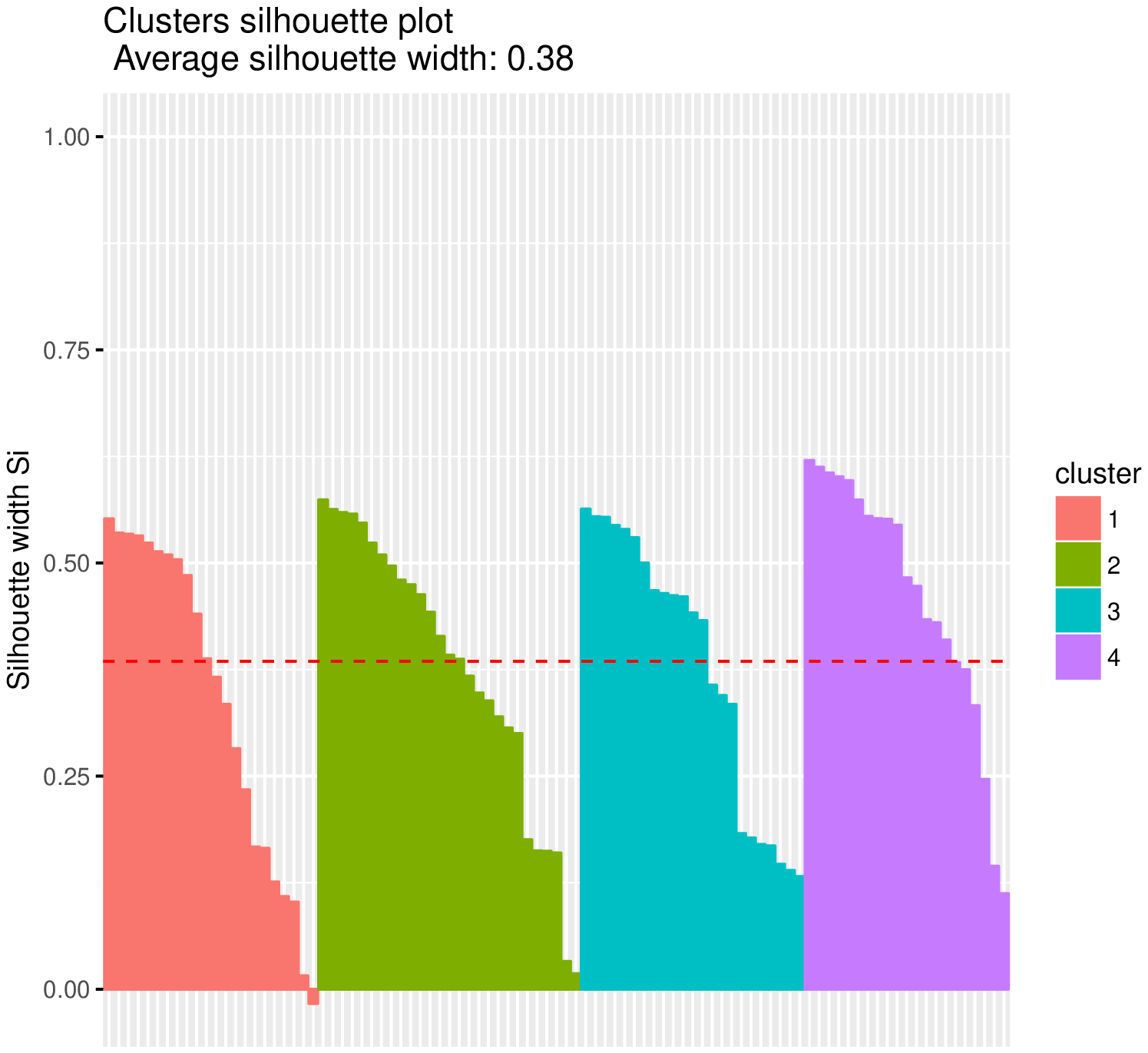}}
\subfloat[][]
   {\includegraphics[width=.33\textwidth, height = 3cm]{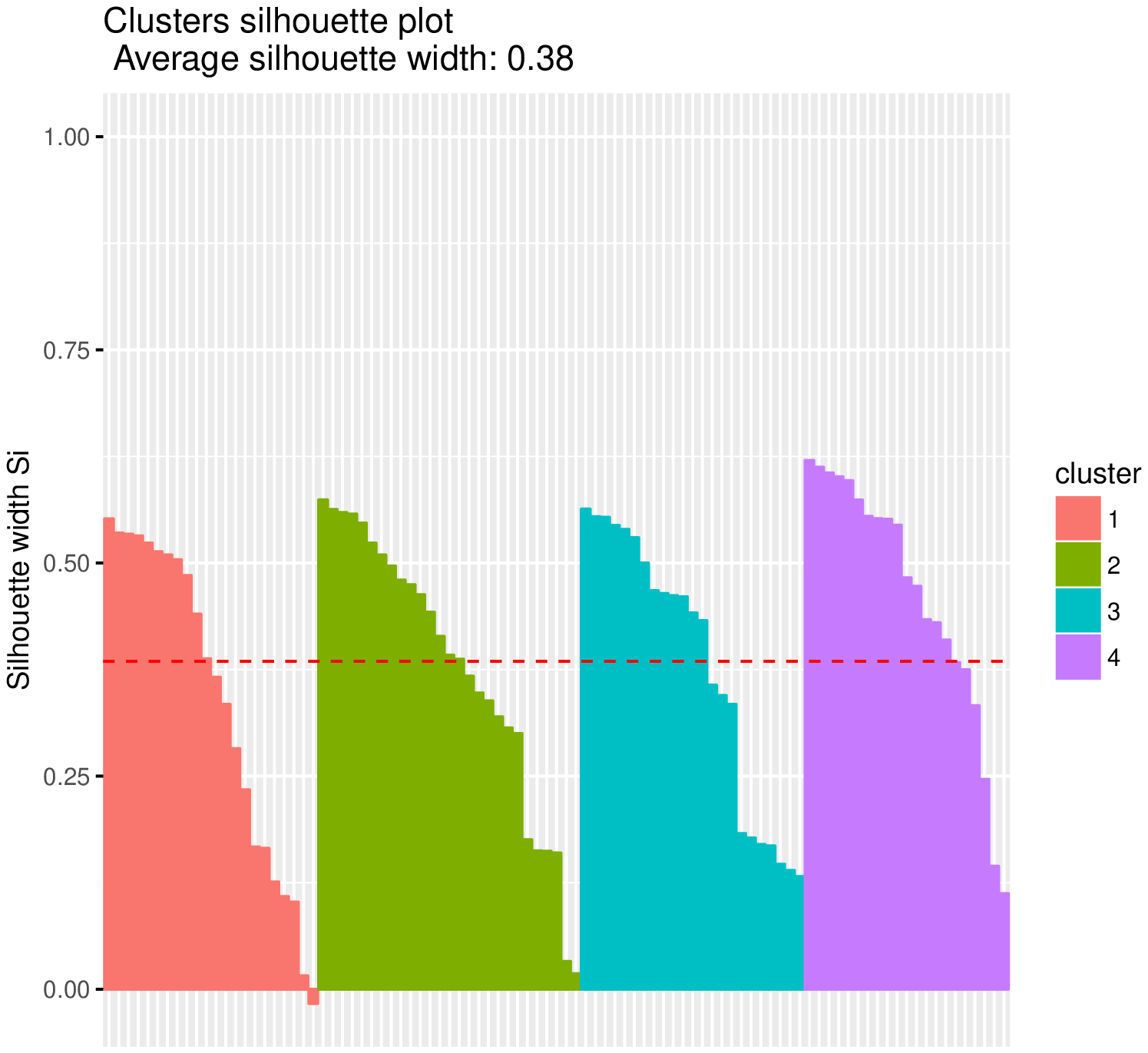}}
\caption[Silhouette plots of the clustering result obtained via the multivariate functional k-means procedure.]{Silhouette plots of the clustering result obtained via the multivariate functional k-means procedure for the Growth dataset, setting (a) k=2, (b) k=3, (c) k=4 and (d) k=5 with distance $\hat{d}_p$ and log$_{10}(p)=8$: the data are ordered according to an increasing value of silhouette within each cluster and the colour indicates the cluster assignment.}
\label{fig:phoneme_sil}
\end{figure}

The $\hat{d}_p$ distance is computed with the eigenvalues $\{\hat{\lambda}_k;\, 1 \leq k \leq T\}$ and the associated eigenfunctions $\{\hat{\varphi}_k;\, 1 \leq k \leq T\}$ derived from the estimated covariance function.
The growth curves are shown in Figure \ref{fig:growth} (a), where they appear very similar and quite indistinguishable from each other; this would suggest from a preliminary analysis that we should study their micro-structure. In Figure \ref{fig:growth} (b) we show the performance of the $k$-means algorithm with the $L^2$-distance (green solid line), the truncated Mahalanobis semi-distance $d_M^K$ with $K=3$ (blue solid line) and the $\hat{d}_p$ distance (black line), along with some numerical results in Table \ref{tbl:growth}. The situation is quite similar to case (ii) of Section \ref{simulation}, where the $k$-means algorithm gives better results only with the $\hat{d}_p$ distance and for high values of $p$. Indeed, in this case, the $k$-means with the $\hat{d}_p$ distance setting a low value of $p$ is able to correctly classify less than $65\%$ of the curves, only a bit more than $d_{L^2}$ and $d_M^K$, while if we set a high value of $p$, the proportion of correctly classified curves is between $87\%$ and $89\%$. In Figure \ref{fig:phoneme_sil} we show at the silhouette plots computed with the $\hat{d}_p$ distance with $p=10^{8}$ and $k\in\{2,3,4,5\}$
number of cluster, which confirms that the best grouping structure is obtained by setting $k^*=2$.\par
As we could expect by looking at the growth curves in Figure \ref{fig:growth} (a), in this case it is better to set a low value of the parameter $p$,
since the curves seem very similar and the difference involves the micro-structure of the functional data.

\begin{table}
\centering
\resizebox{\columnwidth}{!}{\subfloat[][$d_{L^2}$]
{ \begin{tabular}{ccc}
    \toprule
    Cluster & Girls & Boys \\
    \midrule
    1 & 37 & 17  \\
    2 & 16 & 23  \\
    \bottomrule
    \multicolumn{3}{c}{Correct classification: .6452}\\
    \bottomrule
  \end{tabular}}
\subfloat[][$d_M^K$]
{ \begin{tabular}{ccc}
    \toprule
    Cluster & Girls & Boys \\
    \midrule
    1 & 38 & 18  \\
    2 & 16 & 21  \\
    \bottomrule
    \multicolumn{3}{c}{Correct classification: .6344}\\
    \bottomrule
  \end{tabular}}

\subfloat[][$\hat{d}_p$,  log$_{10}(p) = -2$]
{ \begin{tabular}{ccc}
    \toprule
    Cluster & Girls & Boys \\
    \midrule
    1 & 37 & 17  \\
    2 & 16 & 23  \\
    \bottomrule
    \multicolumn{3}{c}{Correct classification: .6452}\\
    \bottomrule
  \end{tabular}}
\subfloat[][$\hat{d}_p$, log$_{10}(p)=8$]
{ \begin{tabular}{ccc}
    \toprule
    Cluster & Girls & Boys \\
    \midrule
    1 & 47 & 5  \\
    2 & 7  & 34 \\
    \bottomrule
    \multicolumn{3}{c}{Correct classification: \textbf{.8710}}\\
    \bottomrule
  \end{tabular}}}
\caption[Confusion matrices related to the functional $k$-means  for the growth curves.]{Confusion matrices related to the functional $k$-means  for the growth curves.}
\label{tbl:growth}
\end{table}

\section{Case study II: ECG dataset}
\label{ECG}
In this section we apply the functional $k$-means algorithm to a real case study on electrocardiographics signals (ECGs).
The dataset provided by Mortara-Rangoni S.r.l. contains ECG signals, which represent a recording of the electrical activity of the heart over a period of time. Each signal consists of 8 curves, such that we have a multivariate functional dataset with $J=8$. \par
Among the signals in the dataset, some are healthy while others are affected by Bundle Branch Blocks. Depending on the anatomical location of the defect which leads to a bundle branch block, the blocks are further classified into right bundle branch block (RBBB) and left bundle branch block (LBBB).
%
%
%
%
The aim of the analysis is to establish if there is statistical evidence of shape modifications induced on the ECG curves by the pathologies.
The investigation will be conducted only from a statistical perspective, without considering any clinical criteria.\par
The ECG signals consist of noisy and discrete observations of the functions describing the ECG traces of the patients. Moreover, each patient has his own 'biological' time, i.e. the same event of the heart dynamics may occur at different times for different patients; that is why the morphological change due to this difference in timings is misleading from a statistical perspective. To address these two problems, which are quite popular in functional data analysis, the data have been previously smoothed and registered; see \cite{Ieva.et.al.13} for further details. \par
We consider $n=700$ subjects, where among them 400 are healthy, 150 are affected by LBBBs and 150 are affected by RBBBs. From the sample covariance function
we estimate the eigenvalues $\{\hat{\lambda}_k;\, k \geq 1\}$ and the associated eigenfunctions $\{\hat{\boldsymbol{\varphi}}_{k}=(\hat{\varphi}_k^{(1)}, \ldots, \hat{\varphi}_k^{(8)})^\top, k\geq 1\}$, which are used to compute the generalized Mahalanobis distance $\hat{d}_p$ as defined in \eqref{def:hat_dp}.
To perform comparisons and to test the robustness of the $k$-means algorithm based on the $\hat{d}_p$ distance,
we have considered as competitors the same distances used in Section~\ref{simulation}, i.e. $d_M^K$ and $d_{L^2}$.

\begin{figure}
\centering
\subfloat[][]
   {\includegraphics[width=.33\textwidth, height = 3cm]{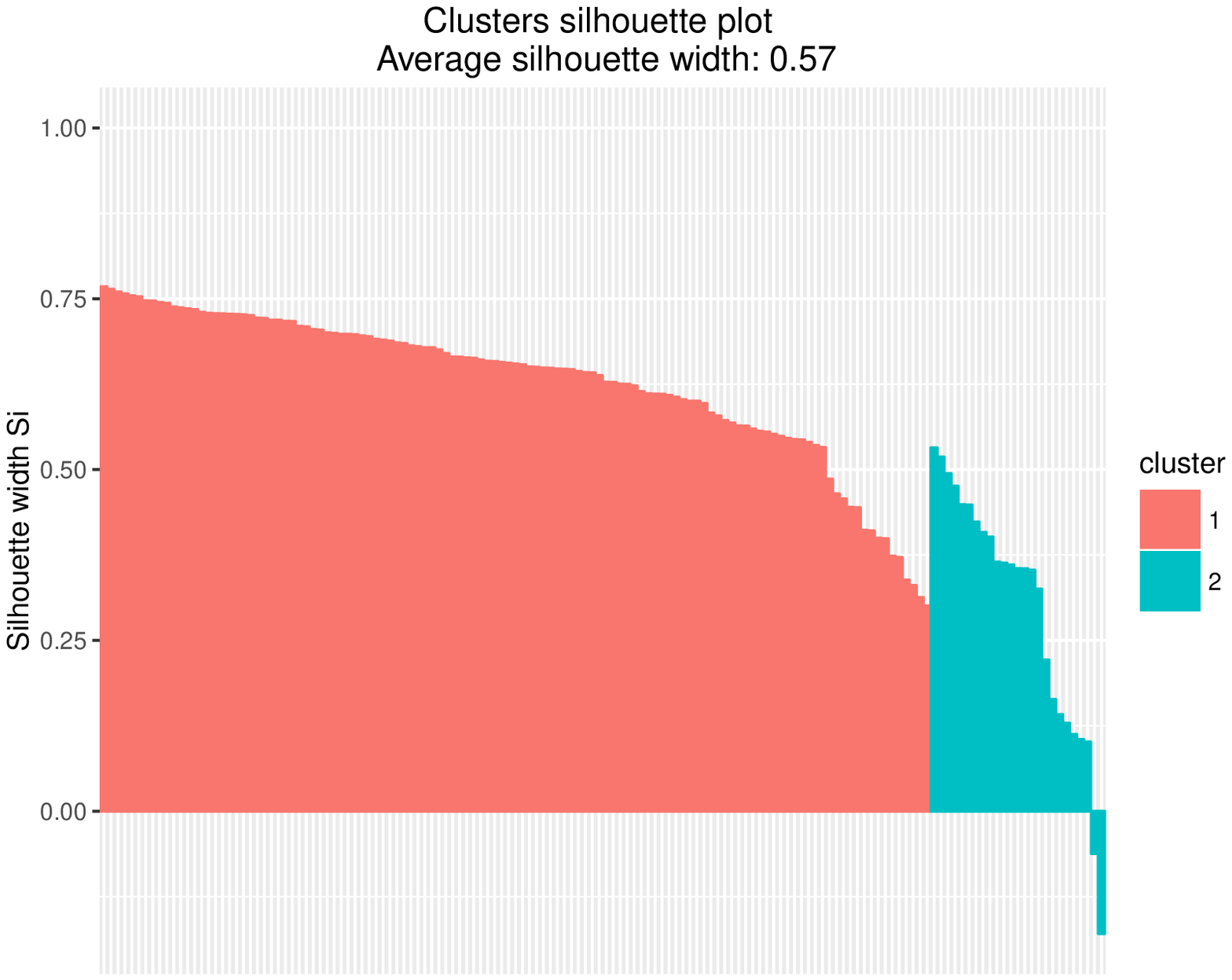}}
\subfloat[][]
   {\includegraphics[width=.33\textwidth, height = 3cm]{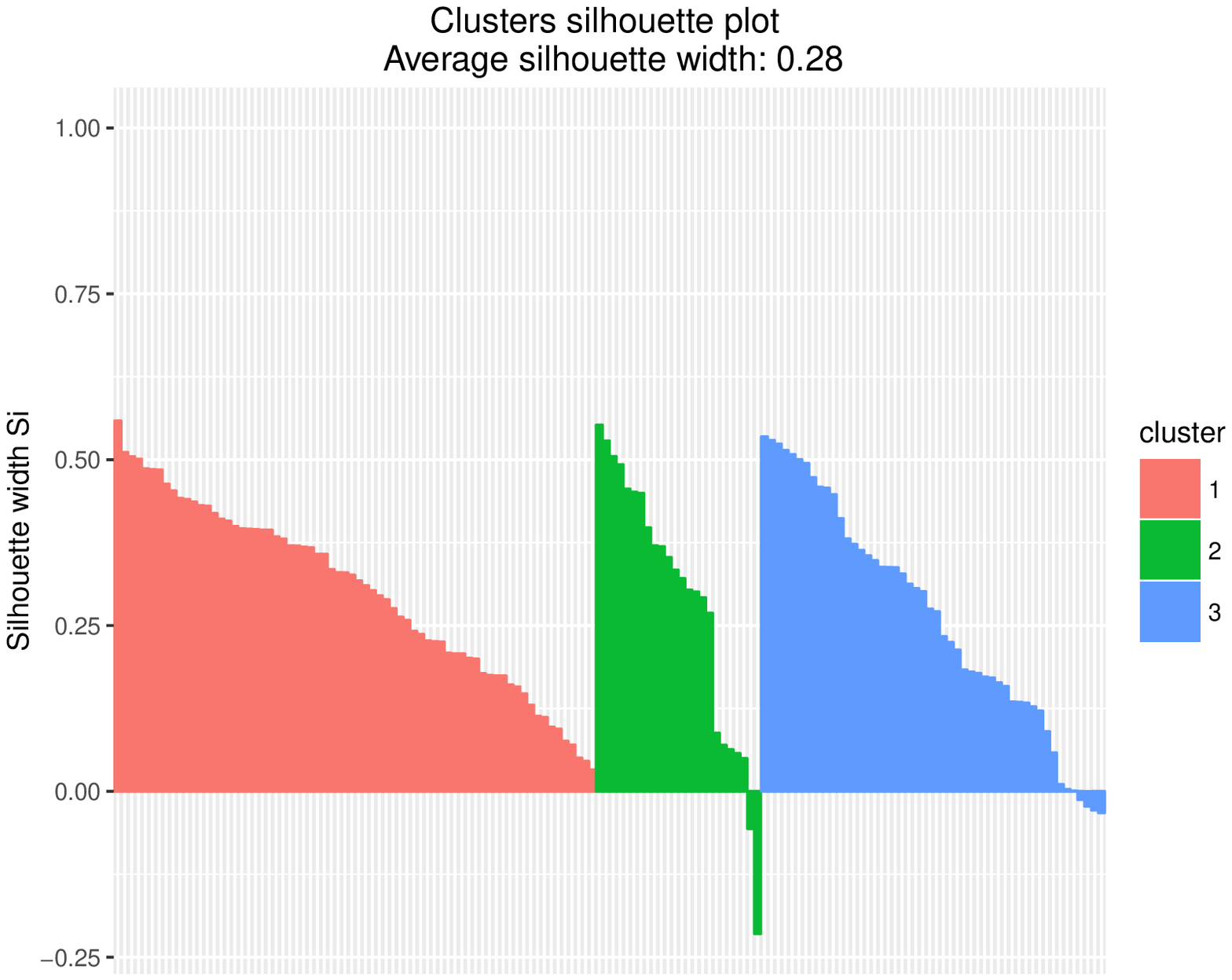}}
   \\
\subfloat[][]
   {\includegraphics[width=.33\textwidth, height = 3cm]{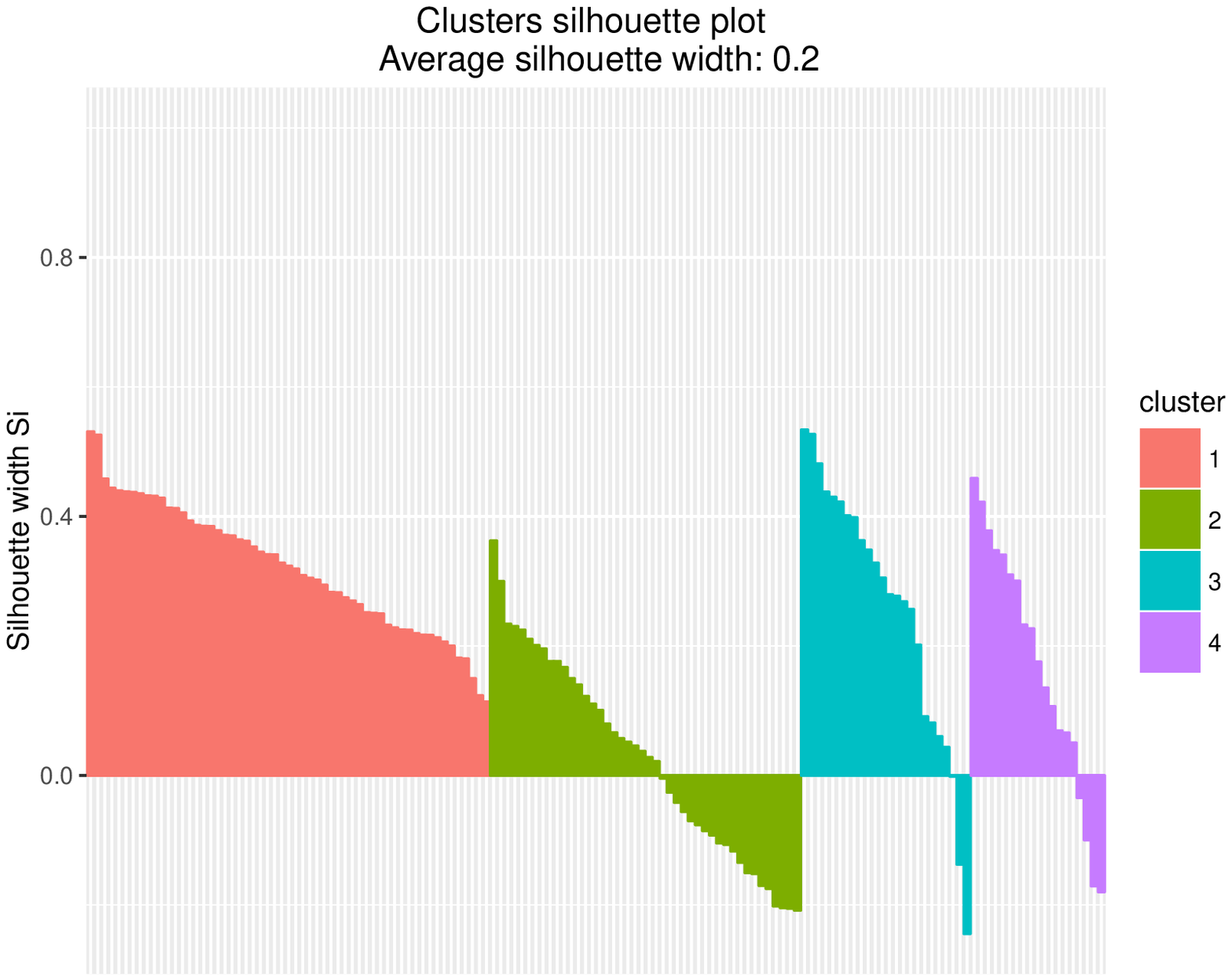}}
\subfloat[][]
   {\includegraphics[width=.33\textwidth, height = 3cm]{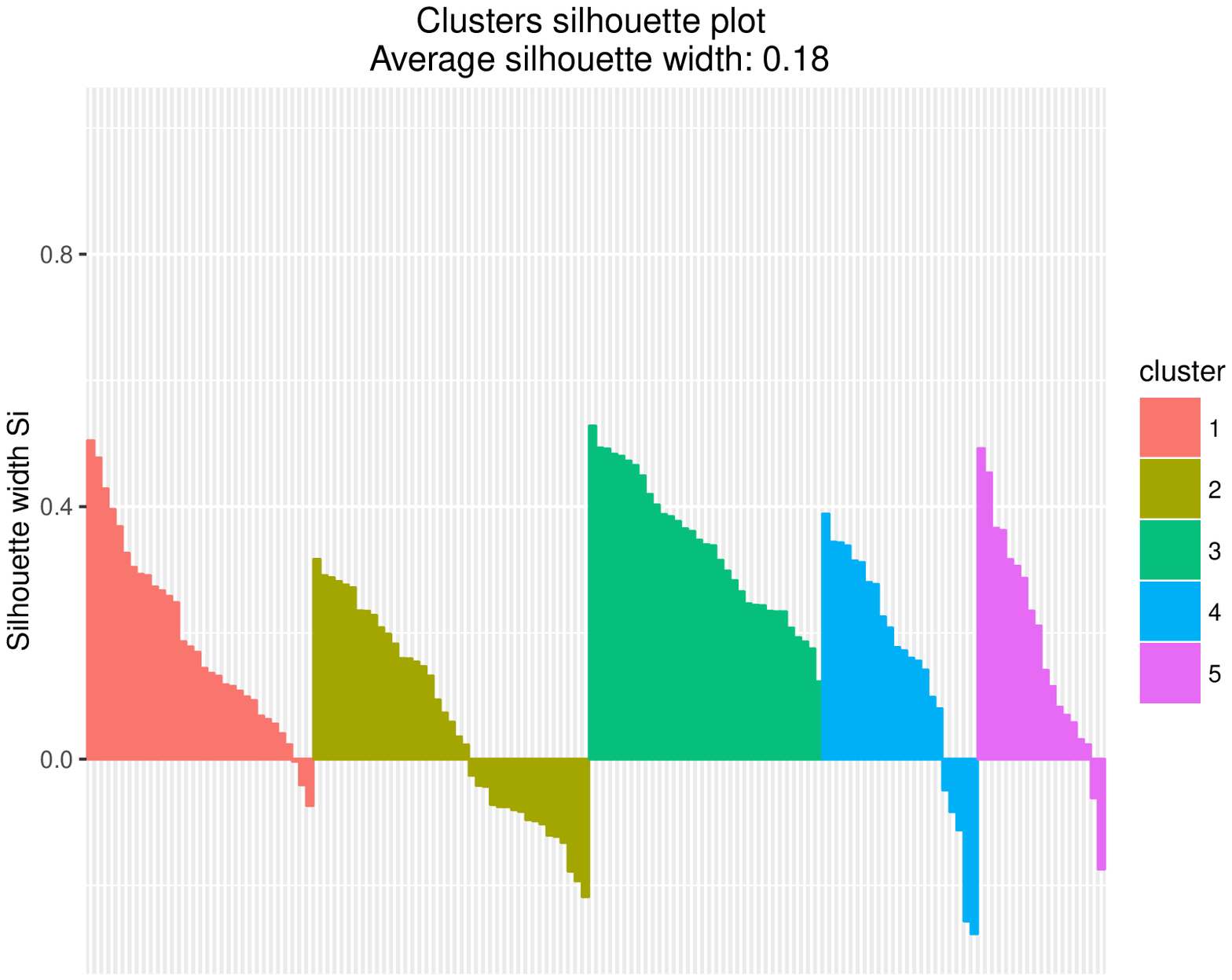}}
\caption[Silhouette plots of the clustering result obtained via the multivariate functional k-means procedure.]{Silhouette plots of the clustering result obtained via the multivariate functional k-means procedure for the ECG dataset, setting (a) k=2, (b) k=3, (c) k=4 and (d) k=5 with distance $\hat{d}_p$ and log$_{10}(p)=-4$: the data are ordered according to an increasing value of silhouette within each cluster and the colour indicates the cluster assignment.}
\label{fig:silhouette}
\end{figure}
Figure \ref{fig:silhouette} shows the final silhouette plots obtained by clustering the multivariate samples of ECG traces according to the functional $k$-means procedure with the $\hat{d}_p$ distance, with $p=10^{-2}$ and $k=\{2,3,4,5\}$. As we can see from the figure, the grouping structure obtained by setting $k=3$ seems the best, both in terms of silhouette profile and wrong assignments. A similar result is obtained by measuring the distance between curves with the $d_M^K$ or the $d_{L^2}$ distances; we thus set $k^*=3$. Moreover, the $k$-means seems to detect the best grouping structure when we use the $\hat{d}_p$ distance with small values of the parameter $p$.\par
Because of the high computational cost due to the construction of the $\hat{d}_p$ distance, that takes into account a large number of components, the code has been parallelized using the R-packages \texttt{doParallel} and \texttt{foreach} (for further details about both packages, see \cite{Weston.et.al.15.2} and \cite{Weston.et.al.15.1}). This has greatly reduced the computational time of the algorithm.
The results obtained by the $k$-means multivariate clustering procedure with all the three distances are shown in the confusion matrices of Table \ref{tbl:ecg}. We a posteriori identify the cluster with the greater number of physiological ECG traces as the one containing the healthy subjects. Subsequently, to distinguish the clusters corresponding to the pathological traces, we first select the cluster containing the maximum number of pathological traces of the same kind and at last the remaining cluster.\par
\begin{table}
\small
\centering
\subfloat[][$L^2$ distance]
{ \begin{tabular}{cccc}
    \toprule
    Cluster & Healthy & LBBB & RBBB \\
    \midrule
    1 & 355 & 18 & 29\\
    2 & 40 & 96 & 1\\
    3 & 5 & 36 & 120\\

    \bottomrule
    \multicolumn{4}{c}{Correct classification: .8228}\\
    \bottomrule
  \end{tabular}}
\subfloat[][$d_M^K$ distance]
{ \begin{tabular}{cccc}
    \toprule
    Cluster & Healthy & LBBB & RBBB \\
    \midrule
    1 & 362 & 24 & 36\\
    2 & 2 & 92 & 1\\
    3 & 36 & 34 & 113\\
    \bottomrule
    \multicolumn{4}{c}{Correct classification: .8142}\\
    \bottomrule
  \end{tabular}}
\\
\subfloat[][$\hat{d}_p$,  log$_{10}(p) = -4$]
{ \begin{tabular}{cccc}
    \toprule
    Cluster & Healthy & LBBB & RBBB \\
    \midrule
    1 & 396 & 28 & 24\\
    2 & 3 & 96 & 0\\
    3 & 1 & 3 & 126\\
    \bottomrule
    \multicolumn{4}{c}{Correct classification: \textbf{.8830}}\\
    \bottomrule
  \end{tabular}}
\subfloat[][$\hat{d}_p$,  log$_{10}(p) = 4$]
{ \begin{tabular}{cccc}
    \toprule
    Cluster & Healthy & LBBB & RBBB \\
    \midrule
    1 & 321 & 44 & 40\\
    2 & 64 & 95 & 14\\
    3 & 15 & 11 & 96\\
    \bottomrule
    \multicolumn{4}{c}{Correct classification: .7314}\\
    \bottomrule
  \end{tabular}}
\caption[Confusion matrices related to the functional $k$-means for the ECG traces.]{Confusion matrices related to the functional $k$-means for the ECG traces.}
\label{tbl:ecg}
\end{table}
\begin{figure}
\centering
   {\includegraphics[width=.5\textwidth]{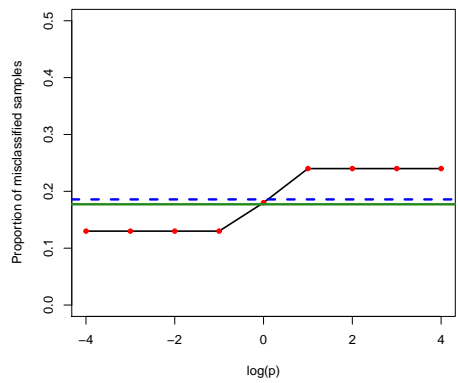}}
\caption[Proportion of misclassified samples with the functional $k$-means for the ECG dataset using the $L^2$ distance, the $d_M^K$ distance and the $\hat{d}_p$ distance.]{Proportion of misclassified samples with the functional $k$-means for the ECG dataset using the $\hat{d}_p$ distance (black line), with the $L^2$-distance (blue dashed line) and with the $d_M^K$ semi-distance (green solid line).}
\label{fig:misclecg}
\end{figure}
\begin{figure}
\centering
\subfloat[][]
   {\includegraphics[width=.4\textwidth, height = 4cm]{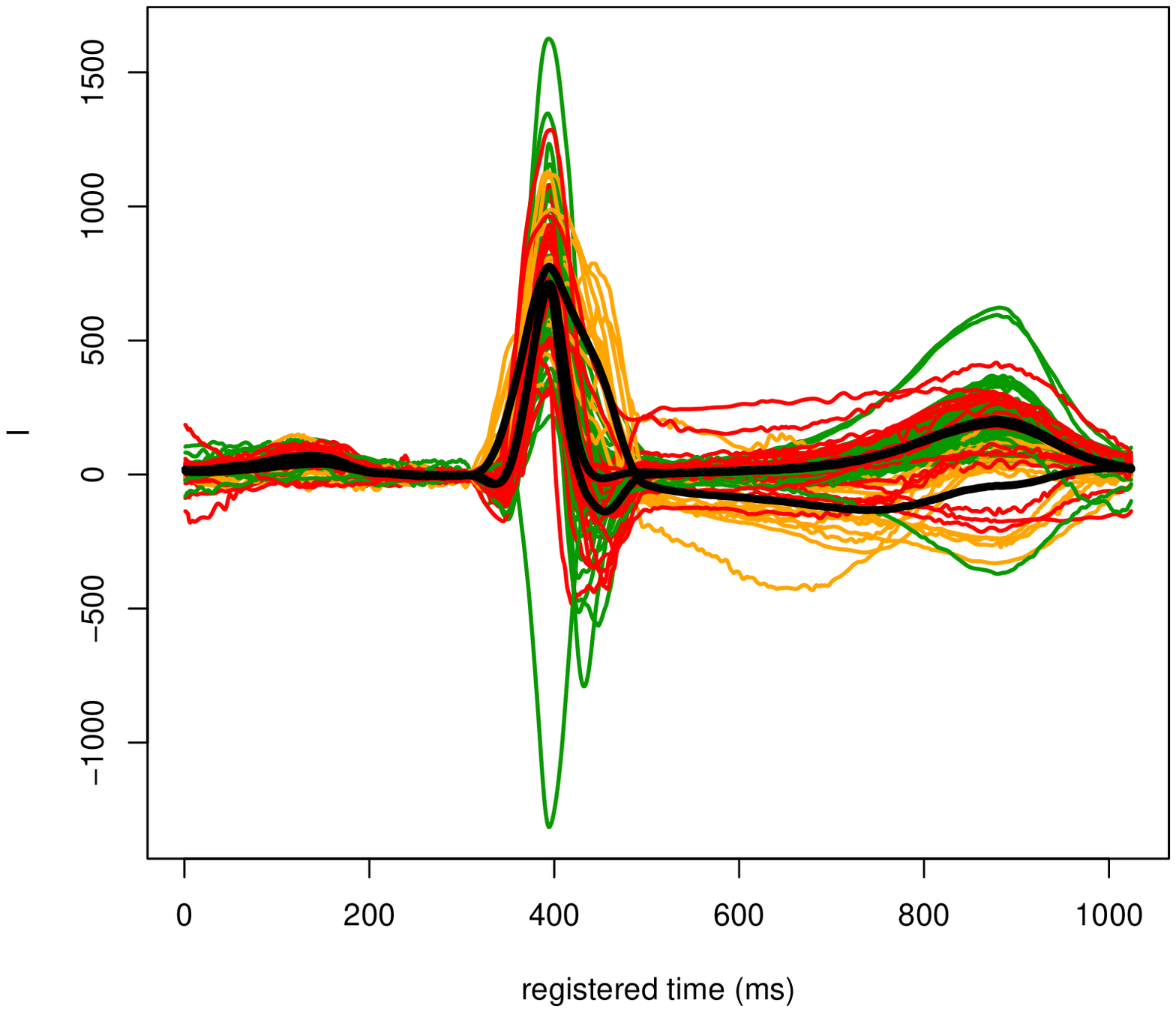}}
\subfloat[][]
   {\includegraphics[width=.4\textwidth, height = 4cm]{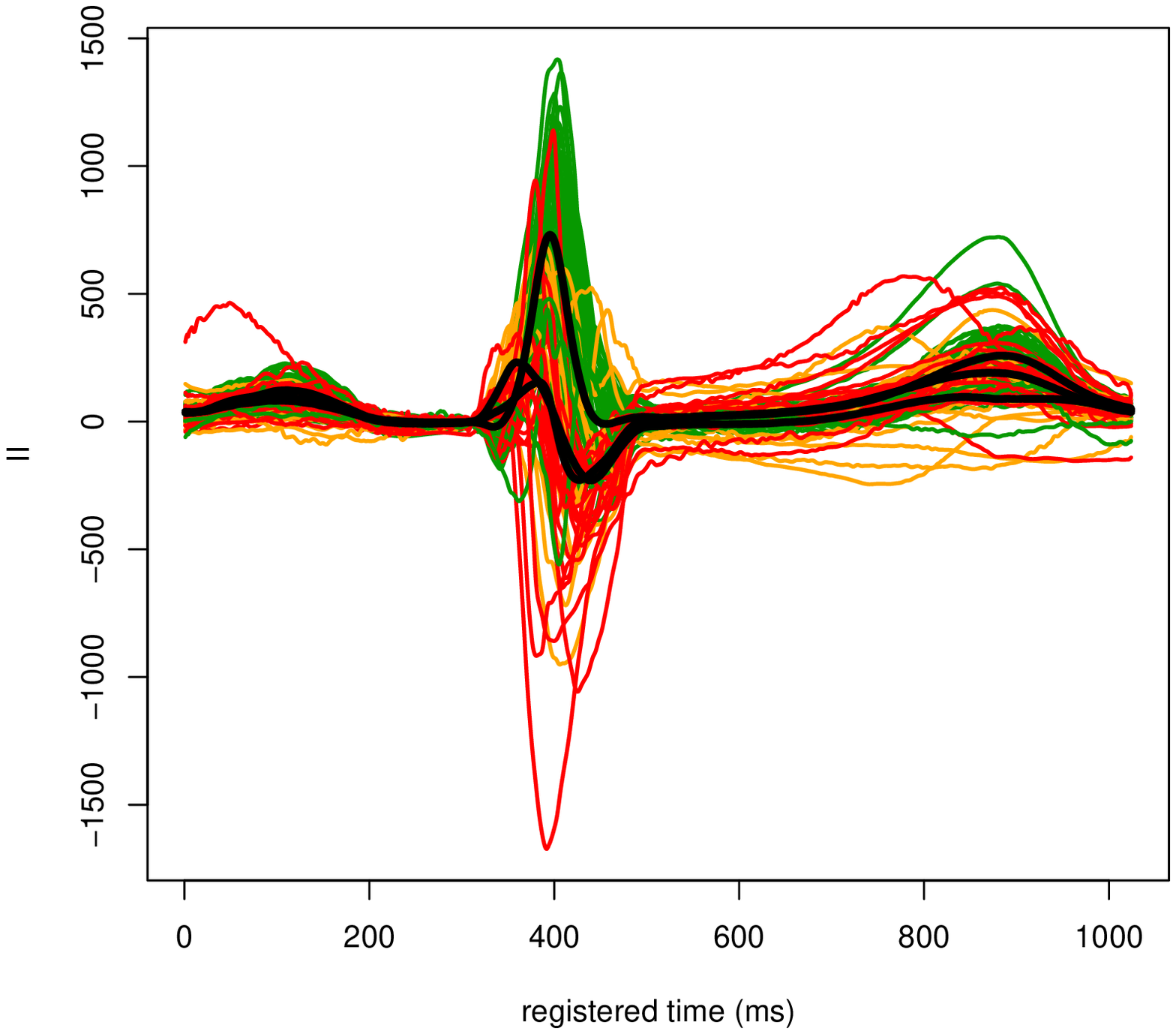}}
%
\caption[ECG leads assigned to each cluster.]{Curves assigned to each cluster in the first two ECG leads (green for the healthy subjects, orange for the LBBBs, red for the RBBBs).}
\label{fig:leads}
\end{figure}
Looking at the four confusion matrices, we can note that the obtained results are quite good and they differ a little depending on the tested distance. As obtained in case (iii) of Section~\ref{simulation}, from Figure \ref{fig:misclecg} we can see that, the higher is the value of the parameter $p$ in the $\hat{d}_p$ distance, the higher is also the number of misclassified curves by the $k$-means. In particular, in this case we go from more than $88\%$ of well-classified subjects to about $73\%$. Then, we can state that, in this case, the generalized Mahalanobis distance with small values of $p$ is the best choice; this performance are even better than those with the $L^2$ distance and the truncated Mahalanobis semi-distance $d_M^K$. As discussed in Section~\ref{simulation}, this scenario can be explained by the fact that the differences among the ECG signals concern the macro-structure of the curves, i.e. differences in the amplitude and inversion of some parts of the curves, which are better identified by the $\hat{d}_p$ distance with low values of $p$.\par
Figure \ref{fig:leads} shows the ECG curves of the subjects considered in this study, in the first two of the 8 leads and with a different color for each cluster (green for the healthy subjects, orange for the LBBBs, red for the RBBBs). Looking at the black centroids in Figure \ref{fig:leads}, it is possible to note the main differences between the healthy subjects and those affected by Bundle Branch Blocks.\par

\section{Discussion and future developments}
\label{conclusion}
In this work we have considered the problem of clustering multivariate curves, proposing a functional $k$-means algorithm based on a suitable generalization of the Mahalanobis distance for Hilbert spaces. It has been shown, both in simulations and in two real case studies, that the performances of this method are definitely higher than those obtained with other distances typically used in functional data analysis.\par
Morever, we have discussed that, when the curves in the sample differ mainly in their macro-structure, as for example the ECG signals where there are differences in the amplitude and the inversion of some parts of the curves, the $k$-means algorithm with the $\hat{d}_p$ distance works very well with low values of the parameter $p$, even better than the $L^2$-distance and the truncated Mahalanobis semi-distance. If instead the curves look indistinguishable, as for example the growth curves where each function grows in a slightly different way than the other ones and this difference involves the micro-structure of the curve, the $k$-means algorithm based on the $\hat{d}_p$ distance with high values of $p$ provides the best results, performing remarkably better than the other considered distances.\par
As future development, it will be interesting to investigate the performances of this distance with other clustering algorithms different from the $k$-means; moreover, since this distance can be extended to more complex spaces, such as the Sobolev space $H^1$, we could improve the clustering procedure by incorporating the information on the derivative of the functional data.


\end{document}